\documentclass[12pt]{article}
\usepackage{amssymb,graphicx,amsmath,array,verbatim}

\makeatletter
\@addtoreset{equation}{section}

\renewcommand{\thefootnote}{\fnsymbol{footnote}}
\makeatother

\newcommand {\beq}{\begin{eqnarray}}
\newcommand {\eeq}{\end{eqnarray}}
\def\p{\partial}

\newcommand{\vs}[1]{\vspace{#1 mm}}
\newcommand{\hs}[1]{\hspace{#1 mm}}
\newcommand{\bpm}{\begin{pmatrix}}
\newcommand{\epm}{\end{pmatrix}}

\newcommand{\R}{\mathbb{R}}
\newcommand{\C}{\mathbb{C}}

\newcommand{\tr}{{\rm Tr}}
\newcommand{\D}{\mathcal D}

\newcommand{\ba}{\left( \begin{array}}
\newcommand{\ea}{\end{array} \right)}
\newcommand{\be}{\begin{equation}}
\newcommand{\ee}{\end{equation}}
\newcommand{\bea}{\begin{eqnarray}}
\newcommand{\eea}{\end{eqnarray}}
\newcommand{\beann}{\begin{eqnarray*}}
\newcommand{\eeann}{\end{eqnarray*}}

\newcommand{\Tr}{{\rm Tr}}

\newcommand{\Z}{\mathbb{Z}}

\usepackage[usenames]{color}

\begin{document}
\thispagestyle{empty}
\begin{flushright}
IFUP-TH/2011-8 \\
KUNS-2332 \\
YGHP-11-43 \\
May, 2011
\end{flushright}
\vspace{3mm}
\begin{center}
{\LARGE \bf 
Dynamics of Non-Abelian Vortices} \\

\vspace{0.5cm}
{\normalsize\bfseries
Minoru Eto${}^{1}$, Toshiaki Fujimori${}^{2}$, 
Muneto Nitta${}^{3}$,\\ Keisuke Ohashi${}^{4}$ and
Norisuke Sakai${}^{5}$}
\footnotetext{
Email addresses: \tt
meto(at)sci.kj.yamagata-u.ac.jp, 
toshiaki.fujimori(at)pi.infn.it, \\
nitta(at)phys-h.keio.ac.jp, 
ohashi(at)gauge.scphys.kyoto-u.ac.jp,
sakai(at)lab.twcu.ac.jp}

\vskip 1.5em
{\it\small 
$^1$ Department of Physics, Yamagata University, Yamagata 990-8560, Japan}\\
{\it\small 
$^2$ INFN, Sezione di Pisa, 
Largo B.~Pontecorvo, 3,  Ed.~C, 56127 Pisa, Italy, and}\\
{\it\small 
Department of Physics,``E. Fermi'', University of Pisa, 
Largo B.\,Pontecorvo,\,3, Ed.\,C, 56127 Pisa, Italy}
\\{\it\small
$^3$ Department of Physics, and Research and Education 
Center for Natural Sciences,}\\  
{\it\small 
 Keio University, Hiyoshi 4-1-1, Yokohama, Kanagawa 223-8521, Japan}\\
{\it\small 
$^4$ Department of Physics, Kyoto University, Kyoto 
606-8502, Japan}\\
{\it\small 
$^5$ 
Department of Mathematics, Tokyo Woman's Christian University, 
Tokyo 167-8585, Japan
}\\
\vspace{10mm}

\begin{abstract}
The scattering is studied using 
moduli space metric for well-separated vortices of 
non-Abelian vortices in (2+1)-dimensional $U(N)$ gauge theories with $N$ 
Higgs fields in the fundamental representation. 
Unlike vortices in the Abelian-Higgs model, 
dynamics of non-Abelian vortices has a lot of new features;
The kinetic energy in real space can be 
transfered to that of internal orientational moduli and vice versa, 
the energy and charge transfer between two vortices, 
the scattering angle of collisions with a fixed impact parameter 
depends on the internal orientations,  
and some resonances appear due to synchronization of the orientations.
Scattering of dyonic non-Abelian vortices 
in a mass deformed theory is also studied. 
We find a bound state of two vortices moving 
along coils around a circle, like a loop of a phone code.
\end{abstract}

\end{center}

\vfill
\newpage
\thispagestyle{empty}
\tableofcontents

\newpage
\setcounter{page}{1}
\setcounter{footnote}{0}
\renewcommand{\thefootnote}{\arabic{footnote}}
\section{Introduction \label{sec:1}}

Topological solitons are localized finite energy 
solutions of classical equations of motion in field theories.
Their stability is protected by topological winding numbers. 
Various topological solitons 
are known so far in field theories, for instance, 
instantons, magnetic monopoles, vortices 
and kinks (domain walls) \cite{Manton:2004tk}. 
They are not only the classical solutions of the 
equations of motion but also play important roles in 
non-perturbative quantum effects through, 
for example, strong-weak dualities such as 
the electromagnetic dual in four dimensions.
While topological solitons have been observed in condensed 
matter systems, 
in cosmology, solitons, especially cosmic strings, 
may be observed in a future by a direct detection of 
gravitational waves, the gravitational lens or 
the cosmic microwave background.

In each dimension, 
we can consider dynamics of various topological solitons 
which behave as if they are particle-like objects 
; kinks in $d=1+1$, 
vortices in $d=2+1$, magnetic monopoles in $d=3+1$, 
and instantons in $d=4+1$. 
In principle, the motion of solitons are determined by field equations 
which are usually highly nonlinear differential equations. 
Therefore a very complicated analysis is required 
to understand the soliton dynamics in general.
For complicated physical systems, 
it is important to extract essential degrees of freedom 
by throwing away other unimportant degrees of freedom. 
This procedure is a challenging and interesting problem. 
Among various kinds of solitons, 
gauge theories often admit an important class of solitons, 
called {\it local solitons};
Yang-Mills instantons \cite{Belavin:1975fg}, 
't Hooft-Polyakov monopoles \cite{'tHooft:1974qc}, 
Abrikosov-Nielsen-Olesen (ANO) vortices \cite{Abrikosov:1956sx} 
and ${\mathbb C}P^1$ kinks (domain walls) 
\cite{Abraham:1992vb,Gauntlett:2000ib}.
A nice way of extracting such essential 
degrees of freedom has been established  
for a particularly important
class of local solitons, 
called Bogomol'nyi-Prasad-Sommerfield (BPS) solitons 
\cite{Bogomolny:1975de}. 
BPS solitons saturate the lower energy bound and are 
the most stable among configurations with a fixed topology 
\cite{Bogomolny:1975de}.
They further naturally appear in supersymmetric gauge theories, 
break/preserve a fraction of supersymmetry, 
and consequently are quantum mechanically stable 
under perturbative or non-perturbative quantum corrections 
\cite{Witten:1978mh}.
Since static BPS solitons exert 
exactly no forces among them at any distance, 
multiple solitons can statically coexist at any position, 
which become parameters of multi-soliton solutions, 
namely moduli parameters associated with massless modes of solitons. 
Even though no static forces exist, solitons feel forces depending on 
their velocity.
As long as solitons move slowly, it is sufficient to 
consider the motion of massless modes to describe the dynamics 
neglecting all the massive modes. 
This approximation is called the moduli space 
(geodesic, or Manton) approximation, 
which was proposed by Manton to discuss the dynamics of 
BPS monopoles \cite{Manton:1981mp,Manton:2004tk}. 
The moduli space dynamics is a good approximation when the 
kinetic energy of solitons is much smaller than any mass scales of the theory.
According to the moduli space dynamics, solitons move 
along geodesics of the moduli space of the soliton, 
where the forces depending on velocities of solitons are represented 
as geodesic forces.

Unfortunately, it is not very easy to get 
the moduli space metric in general. 
Only in few cases, the metrics are known explicitly;  
the prime example is the Atiyah-Hitchin metric of $k=2$ BPS 
monopoles in the $SU(2)$ gauge theory.
For multiple monopoles, 
only asymptotic metrics were obtained 
when monopoles are well separated \cite{Gibbons:1995yw}. 
The moduli spaces of well-separated monopoles 
were also obtained 
in gauge theories with arbitrary gauge groups \cite{Lee:1996kz}. 
The moduli space approximation has been successfully 
applied to many other solitons such as 
ANO vortices in the BPS limit 
\cite{Taubes:1979tm,Ruback:1988ba,Shellard:1988zx,
Samols:1991ne,Manton:2002wb,Chen:2004xu}, 
lumps \cite{Ward:1985ij}, 
and BPS domain walls \cite{Gauntlett:2000ib}. 
It has been applied even to  
BPS composite solitons of different kinds 
\cite{Eto:2006bb,Eto:2008mf}, 
such as domain wall networks (webs) \cite{Eto:2005cp,Eto:2005sw,Eto:2006pg} 
and vortex-strings stretched between 
parallel domain walls \cite{Isozumi:2004vg,Eto:2005sw,Eto:2006pg}. 
However only the asymptotic metric 
is explicitly known for well-separated ANO vortices 
\cite{Samols:1991ne,Manton:2002wb,Chen:2004xu}, 
because the ANO vortex equations are not integrable.\footnote{
For ANO vortices in a hyperbolic plane with a particular curvature, 
the vortex equations become integrable \cite{Witten:1976ck} 
and consequently the moduli space metric can be calculated 
\cite{Krusch:2009tn}.} 
These metrics have been used to discuss the scattering problems.
In particular, scattering processes of solitons have 
attracted attention of many mathematicians and physists. 
It is well known that the magnetic monopoles 
scatter, surprisingly, with 90 degree, when they collide 
head-on \cite{Atiyah:1985dv}. 
The same has been seen for the head-on collision of ANO
vortices in the BPS limit 
\cite{Ruback:1988ba,Shellard:1988zx}, 
and vortex-strings stretched between parallel domain walls \cite{Eto:2008mf}.

In general the moduli space of BPS solitons is a 
surprisingly big space. 
Usually, the dimensions of the moduli space 
is proportional to the topological number 
(the number of the solitons) $k$. 
For example, the ANO vortices 
in the Abelian-Higgs model (Ginzburg-Landau model) 
at the critical coupling have $2k$ moduli parameters, 
which correspond to positions of the vortices 
\cite{Weinberg:1979er}. 
This is much larger than the dimension of the symmetry 
group of the theory. 
Some topological solitons admit more moduli than 
the degrees of freedom of their positions.
For example, the $k$ BPS monopoles in the $SU(2)$ 
gauge theory is known 
to have $4k=(3+1)k$ degrees of freedom. 
As in the vortex case, $3k$ can be 
identified as the monopole positions. 
The remaining $k$ degrees of freedom are 
$U(1)$ phases of the internal space,
which can be called as internal orientations. 
There are other examples of solitons which 
possess the $U(1)$ orientational moduli; 
kinks in the ${\mathbb C}P^1$ model \cite{Abraham:1992vb} 
and $U(1)$ or $U(N)$ gauge theories coupled to 
Higgs fields with non-degenerated masses 
\cite{Gauntlett:2000ib,Isozumi:2004jc}. 
Hence these solitons can be called {\it Abelian solitons}.
On the other hand, non-Abelian moduli are associated with  
{\it non-Abelian solitons}; 
Yang-Mills instantons, 
non-Abelian monopoles \cite{Goddard:1976qe}, 
non-Abelian vortices \cite{Hanany:2003hp}, and 
non-Abelian kinks (in $U(N)$ gauge theory coupled to 
Higgs fields with degenerated masses) 
\cite{Shifman:2003uh}.  
The internal orientations can be regarded as 
Nambu-Goldstone zero modes corresponding to 
global symmetries or (global parts of) local symmetries, 
which are unbroken in the vacuum but 
are spontaneously broken in the presence of solitons.
In general, the motion of internal $U(1)$ orientations 
give preserved charges to the solitons, 
making them BPS dyonic solitons.

Since the discovery of non-Abelian vortices \cite{Hanany:2003hp}, 
much progress has been made in recent years \cite{review,Eto:2005sw,Eto:2006pg}. 
Unlike the ANO vortices in the Abelian-Higgs model, 
the non-Abelian vortices have 
non-Abelian internal orientations 
and associated conserved charges; 
In the case of $U(N)$ gauge 
theory with $N$ flavors of Higgs fields in the 
fundamental representation, 
the internal orientation is the complex projective space 
${\mathbb C}P^{N-1}$,  
which corresponds to Nambu-Goldstone modes associated with 
the $SU(N)_{\rm C+F}$ color-flavor locked global symmetry 
spontaneously broken in the presence of vortices. 
Because of non-Abelian internal orientations, we can expect that the dynamics of the non-Abelian vortices is much richer and more interesting compared to the ANO vortices, 
although the analysis gets much more complicated. 
The moduli space of multiple vortices with full moduli parameters 
was completely determined without metric 
by partially solving BPS vortex equations 
\cite{Isozumi:2004vg,Eto:2005yh,Eto:2006pg,Eto:2006cx};  
The moduli space for $k$ separated vortices is a $k$-symmetric product 
\beq
 {\mathcal M}_k^{\rm sep} 
\simeq ({\mathbb C} \times {\mathbb C}P^{N-1})^k/{\mathcal S}_k 
  \;\;\subset {\mathcal M}_k
\eeq 
of the single vortex moduli space \cite{Eto:2005yh} 
while the whole space ${\mathcal M}_k$ is regular.
General formula for the moduli space metric and its K\"ahler potential 
were given in \cite{Eto:2006uw}.
The metric of the moduli subspace for two coincident vortices 
\cite{Eto:2006db,Eto:2010aj,Eto:2006dx}, 
which is supplement to ${\mathcal M}_{k=2}^{\rm sep}$ inside 
the whole space ${\mathcal M}_{k=2}$,  
was found, 
and it surprisingly shows that 
two non-Abelian vortices scatter with 90 degree 
in head-on collision even though they have different internal orientations 
${\mathbb C}P^{N-1}$ as the initial conditions \cite{Eto:2006db}. 
Most recently, we have obtained the asymptotic metric 
on the moduli space ${\cal M}_k^{\rm sep}$ of $k$ well-separated 
non-Abelian vortices which is valid when the 
separation of vortices are much larger than 
the inverse Compton wave length of massive vector bosons, 
which is the length scale of the vortices \cite{Fujimori:2010fk}.

In this paper, we study the dynamics 
of non-Abelian vortices by using the recently found 
asymptotic metric of non-Abelian vortices 
in the $U(N)$ gauge theory with $N$ Higgs scalar fields 
in the fundamental representation. 
In order to solve the dynamics, we will make use of 
the technique of the moduli space approximation.
The asymptotic metric of the moduli space 
allows us to solve the dynamics of the well-separated 
and slowly moving non-Abelian vortices. 
We find that the dynamics of the non-Abelian vortex 
is quite different from that of the Abelian one. 
The major reason of the difference can be traced 
back to the conserved charges which are absent 
in the Abelian case. 
The vortices with the same charges repel while 
those with the opposite charges attract. 
Since the charges can change during the scattering 
process, non-Abelian vortices experience rich 
and subtle forces, which produce sometimes 
the counter-intuitive or unexpected dynamics. 
We find several new features of the dynamics of non-Abelian 
vortices: 
i) the scattering angle depends on the internal orientation, 
especially parallel orientations give repulsion 
while anti-parallel orientations give attraction,
ii) the energy of real and internal spaces can be transfered,
iii) the energy and charge transfer between two vortices occur, 
and 
iv) some resonances appears due to synchronization of the orientations.

We also study the dynamics of the dyonic 
non-Abelian vortices in the mass deformed theory 
\cite{Collie:2008za}, with the method of the 
moduli space dynamics. 
A new feature in this case is that a potential term 
appears in the low energy effective action. 
Therefore, the solitons experience two 
forces: The one is the geometric force and the other 
is the potential force. 
In a special situation, the two dyonic vortices 
drift on a circular orbit. 
This fact strongly suggests that the non-Abelian 
dyonic vortices can have a bound state.

The paper is organized as follows. 
In Section~\ref{sec:2} we introduce the model which allows the BPS non-Abelian 
vortices and review briefly the low energy effective 
theory of the two non-Abelian vortices, and 
the moduli space dynamics. 
We also give the asymptotic metric on the 
moduli space and the geodesic equations for the well-separated 
vortices. 
In Sec.~\ref{sec:3}, we study the scattering of 
two non-Abelian vortices. 
After defining the Noether charges of vortices, 
we give typical examples of numerical solutions 
of geodesic motion on the moduli space in Subsec.~\ref{subsec:numerical}.
We then analytically study properties of dynamics;
the geodesic force in Subsec.~\ref{subsec:force}, 
scattering of two vortices with a large impact parameter 
by free motion approximation in Subsec.~\ref{subsec:large}, 
and dynamics with zero impact parameter in Subsec.~\ref{subsec:zero}.
In Sec.~\ref{sec:4}, we consider a mass deformation 
of the theory and investigate the dynamics 
of the dyonic non-Abelian vortices. 
We find a bound state of two dyonic vortices moving 
along coils around a circle.
Sec.~\ref{sec:5} is devoted for conclusion and discussion.
In Appendix A, the effective action of vortices 
is written in the $U(2)$ case in terms of unit three-component vectors.

\section{Asymptotic metric for non-Abelian vortices \label{sec:2}}

\subsection{Lagrangian and BPS equations}\label{subsec:Lagrangian}

We consider a $U(N)$ gauge theory in $(2+1)$-dimensional 
spacetime with gauge fields $w_\mu$ for $U(1)$ and 
$W_\mu^a~(a=1,\ldots,N^2-1)$ for $SU(N)_C$, which couple to 
$N$ Higgs fields $H^A~(A=1,\ldots,N)$ in the fundamental 
representation of the $SU(N)_C$ gauge group. 
The Lagrangian is given by 
\beq
\mathcal L &=& - \frac{1}{4e^2} (f_{\mu \nu})^2 
- \frac{1}{4g^2} (F^a_{\mu \nu})^2 
+ (\D^\mu H^A)^\dagger \D_\mu H^A - V, \label{eq:L} \\
V &=& \frac{e^2}{2} ( H^\dagger_A t^0 H^A - \xi )^2 
+ \frac{g^2}{2} (H^\dagger_A t^a H^A)^2, 
\eeq
where $\xi$ is the Fayet-Iliopoulos parameter, $e$ and $g$ 
are gauge coupling constants for $U(1)$ and $SU(N)_C$, 
respectively. 
The overall scalar coupling constants in the potential 
$V$ are chosen to be equal to the square of the gauge 
coupling constants, so that the model admits the BPS 
non-Abelian vortices. Thus the model has the three 
coupling constants $e,g,\xi$. In the three dimensional spacetime,  
all of the mass dimensions of $e^2,g^2,\xi$
are unity. 
Our convention is 
$\D_\mu H^A = (\p_\mu + i w_\mu t^0 + i W_\mu^a t^a) H^A$ 
and $f_{\mu \nu} t^0 + F_{\mu \nu}^a t^a = -i [\D_\mu, \D_\nu]$. 
The matrices $t^0$ and $t^a$ are the generators of 
$U(1)$ and $SU(N)_C$, normalized as 
\beq
t^0 = \frac{1}{\sqrt{2N}} \mathbf 1_N, \hs{10} 
\tr ( t^a t^b ) = \frac{1}{2} \delta^{ab}.
\eeq
As is well known, the Lagrangian Eq.\,\eqref{eq:L} can be 
embedded into a supersymmetric theory with eight supercharges. 
The Higgs fields can also be expressed as an $N$-by-$N$ 
matrix on which the $SU(N)_C$ gauge transformations act 
from the left and the $SU(N)_F$ flavor 
symmetry acts from the right 
\beq
H \rightarrow U_C H U_F^\dagger, \hs{10} 
U_C \in SU(N)_C, \hs{5} U_F \in SU(N)_F.
\eeq
The vacuum of this model ($V=0$) is in an $SU(N)_{C+F}$  
color-flavor locking phase ($U_F=U_C$), where the vacuum expectation 
values (VEVs) of the Higgs fields are 
\beq
H = \sqrt{c} \, \mathbf 1_N, \hs{10} 
c \equiv \left(\frac{2}{N}\right)^{1/2} \xi. 
\label{eq:vac}
\eeq
In this vacuum, we have a mass gap with the mass $m_e$ 
for singlet and $m_g$ for adjoint representations of 
$SU(N)_{C+F}$ 
\beq
m_e = e \sqrt{c}, \hs{10} m_g = g \sqrt{c}. 
\eeq 

The energy density for a static configuration 
can be rewritten as
\beq
\mathcal E &=& \frac{1}{2e^2} \tr \left[ f_{12} t^0 - e^2 ( H_A^\dagger t^0 H^A - \xi ) \right]^2 + \frac{1}{2g^2} \tr \left[ F_{12}^a t^a - g^2 ( H_A^\dagger t^a H^A ) \right]^2 \notag \\
&{}& + 4 | \D_{\bar z} H^A |^2 - \xi \, \tr [ f_{12} t^0 ] - i \epsilon^{ij} \p_i (H_A^\dagger \D_j H^A). \phantom{\bigg[}
\label{eq:vortex_density}
\eeq
For configurations with vorticity $k$ (vortex number),
the energy of is bounded from below 
by the following BPS bound 
\beq
E \ \geq \ k M_{\rm v} \equiv - \xi \int d^2 x \, \Tr [ f_{12} t^0 ] \ 
= \ 2 \pi c k, \hs{10} k \in \Z, 
\eeq
where we have assume that the last term 
in Eq.\,\eqref{eq:vortex_density} vanishes at infinity. 
This bound is saturated if the following BPS equations are satisfied: 
\beq
\D_{\bar z} H = 0 , \hs{15} \frac{2}{e^2} f_{12} t^0 
+ \frac{2}{g^2} F_{12}^a t^a  = H H^\dagger - c \mathbf 1_N, 
\label{eq:BPS} 
\eeq 
where $z=x^1+ix^2$ is a complex coordinate. 
One can easily verify that all the solutions of the BPS equations solve the
original equations of motion of the Lagrangian in Eq.\,(\ref{eq:L}).
The integration constants 
(the moduli parameters or the collective coordinates)
contained in the solutions of the BPS equations 
parameterize the set of configurations with degenerate energy, 
that is, the moduli space of BPS vortices $\mathcal M_k$. 
There are $N$ complex moduli parameters for each vortex: 
one of $N$ is position $z_I$ and 
the rest $N-1$ are internal
orientations $\vec \beta_I$ $(I=1,2,\cdots,k)$. 
Since no net forces are exerted among static vortices, 
each vortex has the position moduli $z_I$ as its degree of freedom. 
The internal orientation $\vec \beta_I$ is associated 
with the $SU(N)_{C+F}$ color-flavor symmetry, 
broken by each vortex down to $SU(N-1) \times U(1)$. 
The Nambu-Goldstone zero modes localize on each vortex 
and the corresponding moduli $\vec \beta_I$ parameterize the coset 
\beq
\frac{SU(N)}{SU(N-1) \times U(1)} \cong \C P^{N-1}.
\eeq
In the following, the $(N-1)$-dimensional vector $\vec \beta_I$ 
denotes the inhomogeneous coordinates of $\C P^{N-1}$ 
for the internal orientation of $I$-th vortex. 
The moduli space $\mathcal M_k$ is a $k N$-dimensional K\"ahler manifold 
parameterized by the holomorphic coordinates $z_I$ and $\vec \beta_I$ 
and has the $SU(N)$ isometry acting on $\vec \beta_I$,
which descends from the $SU(N)_{C+F}$ global symmetry in the vacuum.

\subsection{Asymptotic metric for non-Abelian vortices}

The low-energy dynamics of the vortex system can be described by 
an effective Lagrangian in which the moduli parameters 
are promoted to dynamical variables.
The effective Lagrangian for these moduli parameters 
is given in terms of the K\"ahler metric on the moduli space
\beq
L ~=~ g_{i \bar j} \dot \phi^i \dot{\bar \phi}{}^j, \hs{10} (i,j = 1, \cdots, {\rm dim}_{\C} \mathcal \, {\cal M}_k=kN),
\label{eq:Leff}
\eeq
where $\phi^i$ are the holomorphic coordinates on the moduli space:
\beq
\{\phi^i\} = \{z_I, \vec \beta_I\}.
\eeq
The metric $g_{i \bar j}$ of the moduli space consists of 
the free part and the interaction part 
which are given in terms of the corresponding K\"ahler potentials
\beq
g_{i \bar j} ~\equiv~ \frac{\p^2 }{\p \phi^i \p \bar \phi^j} K_{\rm free}
+ \frac{\p^2}{\p \phi^i \p \bar \phi^j} K_{\rm int}. 
\eeq
The free part describes the dynamics of completely isolated vortices\footnote{
The K\"ahler class $4\pi/g^2$ can be determined 
\cite{Shifman:2004dr} from the fact that 
sigma model instantons inside a vortex worldsheet 
are Yang-Mills instantons from the bulk point of view \cite{Eto:2004rz}.
}
\beq
K_{\rm free} = \sum_{I=1}^k \left[ \frac{1}{2} M_{\rm v} |z_I|^2 
+ \frac{4\pi}{g^2} \log (1 + |\vec \beta_I|^2) \right], 
\eeq
where $M_{\rm v} = 2\pi c$ is the tension of the vortex and $4\pi/g^2$ corresponds to
the radius of $\mathbb{C}P^{N-1}$.
On the other hand, the interaction part of the K\"ahler potential describes 
the leading interactions between well-separated vortices \cite{Fujimori:2010fk}
\beq
K_{\rm int} ~\approx~ \sum_{I<J} K^{(I,J)}, 
\quad 
K^{(I,J)} \equiv 
- 2\pi N \left[ \frac{c_e^2}{e^2} K_0(m_e|z_{IJ}|) 
+ \frac{c_g^2}{g^2} \Theta_{IJ} K_0(m_g|z_{IJ}|) \right],
\eeq
where $K_0$ stands for the modified Bessel function of the second kind.
The interaction term $K^{(I,J)}$ 
between $I$-th and $J$-th vortices is 
a function of the relative distance and 
an $SU(N)$ invariant quantity $\Theta_{IJ}$ defined, respectively, by 
\beq
|z_{IJ}| \equiv |z_I - z_J|, \hs{10} \Theta_{IJ} ~\equiv~ 
N \frac{|1 + \vec \beta_I^\dagger \cdot \vec \beta_J|^2}
{(1+|\vec \beta_I|^2)(1+|\vec \beta_J|^2)} - 1. 
\eeq
The origin of the modified Bessel function $K_0$ 
appearing in the interaction term can be traced back to
the asymptotic tail of a profile function of the vortex.
Since the leading term in the modified Bessel function 
$K_0(m|z_{IJ}|)$ is of order $e^{- m |z_{IJ}|}$, 
the interactions exponentially vanish for large $|z_{IJ}|$. 
Note that the mass scales $m_e$ and $m_g$ can be interpreted as 
inverse widths of ``Abelian core" and ``non-Abelian core" of a vortex, 
respectively. 
The strength of the asymptotic coupling is 
controlled by the constants $c_e$ and $c_g$,
which depend on the ratio $m_g/m_e$ and $N$ \cite{Eto:2009wq}. 
The numerical values of $c_e$ and $c_g$ for $U(2)$ vortices 
are given in Table \ref{tab:values}.
{\renewcommand{\arraystretch}{0.75}
\begin{table}[t]
\begin{center}
\begin{tabular}{c|ccccc}
$m_g/m_e$ & $c_e$ & $c_g$ \\
\hline
& & \vs{-3} \\
$0$ & & 1.1363(7) \\
$0.25$ & & 1.1853(1) \\
$0.5$ & & 1.3090(5) \\
$0.75$ & 2.1955(9) & 1.48517(9) \\
$1$ & 1.7078(6) & 1.7078(6) \\
$1.5$ & 1.4714(7) & 2.3031(0) \\
$2$ & 1.4036(9) & 3.14(5) \\
$2.5$ & 1.3746(1) & 4.31(8) \\
$3$ & 1.3594(3) & 5.9(5)  \\
$\infty$ & 1.3266(7) & 
\end{tabular}
\caption{Numerical data for $k=1$ $U(2)$ vortex. 
The constants $c_e$ and $c_g$ are not well-defined 
for $m_e < 2m_g$ and $m_g \rightarrow \infty$, respectively \cite{Eto:2009wq}.}
\label{tab:values}
\end{center}
\end{table}}

For later convenience, let us factor out $4\pi/g^2$ 
from the K\"ahler potential 
and rewrite it in terms of three coupling constants $g,m_e$ and $m_g$ 
instead of $e,g,c$ 
\beq
K_{\rm free} &=& \frac{4\pi}{g^2} \sum_{I=1}^k \left[ \frac{1}{4} m_g^2 |z_I|^2 
+ \log (1 + |\vec \beta_I|^2) \right],
\label{eq:freeKahlerpotential}\\
K_{\rm int} &\approx& \frac{4 \pi}{g^2} \sum_{I<J} \left[
- \frac{N}{2} m_g^2 \left( \frac{c_e^2}{m_e^2} K_0(m_e|z_{IJ}|) 
+ \frac{c_g^2}{m_g^2} \Theta_{IJ} K_0(m_g|z_{IJ}|) \right) \right].
\label{eq:Kahlerpotential}
\eeq

\subsection{Equation of motion along geodesics}

The overall coefficient $1/g^2$ of the effective Lagrangian 
plays a role of the Planck constant 
(loop counting parameter) for fixed mass scales $m_g$ and $m_e$. 
Therefore, the quantum effects are negligible 
and the classical analysis of the effective Lagrangian is valid 
for the energy scale $E$ much larger than $g^2$ 
\beq
g^2 \ll E \ll m_g,\,m_e.
\eeq
Note that the energy scale $E$ should be much smaller than mass gaps 
to justify the use of the low-energy effective Lagrangian. 
In the classical analysis, the dynamics are completely 
independent of the gauge coupling $g$. 

The equations of motion for the moduli parameters $\phi^i$ 
take the following form of the geodesic equation
\beq
\ddot \phi^i + \Gamma^i_{jk} \dot \phi^j \dot \phi^k = 0, \hs{10} \Gamma^i_{jk} \equiv g^{\bar l i} \p_j g_{k \bar l}.
\label{eq:geodesic_eq}
\eeq
Since the interactions are sufficiently small 
($K^{(I,J)} \approx e^{-m |z_{IJ}|}$) for well-separated vortices,
the Christoffel symbol $\Gamma^i_{jk}$ can be approximated as
\beq
\Gamma^i_{jk} ~\approx~ \hat \Gamma_{jk}^i + \hat g^{\bar l i} \hat \nabla_j \p_k \p_{\bar l} K_{int},
\eeq
where $\hat \Gamma^i_{jk}$, $\hat g_{i \bar j}$ and $\hat \nabla_i$ 
are the free part of the Christoffel symbol, metric and covariant derivative respectively.
Then, the equations of motion for $z_I$ can be written as
\beq
\ddot z_I &=& \frac{2}{m_g^2} \sum_{J \not = I} \frac{\p}{\p \bar z_{IJ}} \delta_{IJ} K^{(I,J)},
\label{eq:EOMposition} 
\eeq
where we have defined a differential operator $\delta_{IJ}$ by
\beq
\delta_{IJ} ~\equiv~
\dot z_{IJ}^2 \frac{\p^2}{\p z_{IJ}^2} + 2 \dot z_{IJ} \frac{\p}{\p z_{IJ}}
\left( \dot{\vec \beta}_I \cdot \frac{\p}{\p \vec \beta_I} +  \dot{\vec \beta}_J \cdot \frac{\p}{\p \vec \beta_J} \right) 
+ 2 \dot{\vec \beta}_I \cdot \frac{\p}{\p \vec \beta_I} \ \dot{\vec \beta}_J \cdot \frac{\p}{\p \vec \beta_J} . 
\eeq
The equations of motion for the orientations are
\beq
\frac{\hat \nabla_t \dot{\vec \beta}_I}{1+|\vec \beta_I|^2} 
&=&\sum_{J \not = I} \left[ \left( \frac{\p}{\p \vec \beta_I^\dagger} 
+ {\vec \beta}_I \, \vec \beta_I^\dagger \cdot \frac{\p}{\p \vec \beta_I^\dagger} \right) \delta_{IJ} 
- \frac{2 \dot{\vec \beta}_I}{1+|\vec \beta_I|^2} \dot{\vec \beta}_I \cdot \frac{\p}{\p \vec \beta_I} \right] K^{(I,J)},
\label{eq:EOMorientation}
\eeq
where $\hat \nabla_t \dot{\vec \beta}_I$ is the free part of 
the covariant derivative along the trajectory on $\C P^{N-1}$
\beq
\hat \nabla_t \dot{\vec \beta}_I ~\equiv~ \ddot{\vec \beta}_I 
- \frac{2}{1+|\vec \beta_I|^2} (\vec \beta_I^\dagger \cdot \dot{\vec \beta}_I) \ \dot{\vec \beta}_I. 
\eeq
We refer the right-hand sides of the equations of motion 
(\ref{eq:EOMposition}) and (\ref{eq:EOMorientation}) as forces, 
more precisely, geodesic forces.
In general, they are proportional to squares of velocities $\propto \p_t \phi^i \p_t \phi^j$.


\section{Scattering of $U(2)$ non-Abelian vortices \label{sec:3}}

In this section, we discuss the asymptotic scattering 
of the non-Abelian vortices in the simplest example of the $U(2)$ case 
which shows essential differences between 
the Abelian and the non-Abelian vortices.
In this case, the orientational moduli space becomes 
a sphere $S^2 \simeq \mathbb{C} P^1$.
The inhomogeneous coordinate $\beta_I$ is given 
by the stereographic projection from the north pole of $S^2$ 
and related to the standard spherical coordinates as 
\beq
\beta_I = \tan \frac{\theta_I}{2} e^{i \varphi_I}.
\eeq
One can also make use of the following three-dimensional 
unit vector as coordinates of $S^2$ 
\beq
\vec n_I = ( \sin \theta_I \cos \varphi_I \,,\, 
\sin \theta_I \sin \varphi_I \,,\, \cos \theta_I).
\eeq
Since $SO(3)\simeq SU(2)/\mathbb Z_2$ symmetry is manifest on $\vec n_I$,
the dynamics of the orientational modes 
can be better understood in terms of $\vec n_I$.
The Lagrangian written in terms of $\vec n_I$ is given 
in Appendix \ref{eq:fullLagrangian}.

\subsection{Noether charge} 

One of the sharp contrast of the dynamics of 
non-Abelian vortex to that of Abelian vortex comes from 
the internal orientations.
As mentioned in the introduction, 
such internal degrees of freedom leads to conserved 
Noether charges. 
So let us begin with describing the conserved
charges of non-Abelian vortices.

Corresponding to the $SU(2)$ global symmetry, 
we have one set of conserved charges $\vec Q$. 
In the original theory, they are given by
\beq
\vec Q ~\equiv~ \frac{i}{2} \int d^2 x \, 
\tr \Big[ H \, \vec \sigma \, \D_0 H^\dagger 
- \D_0 H \, \vec \sigma \, H^\dagger \Big],
\eeq
where $\vec \sigma$ are the Pauli matrices. 
Although $\vec Q = 0$ for static non-Abelian vortices, 
these charges arise from the motion of the orientational modes. 

\begin{figure}
\begin{center}
\includegraphics[height=4.5cm]{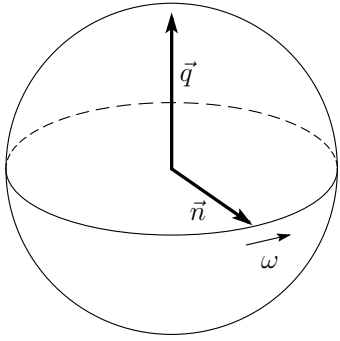}
\end{center}
\caption{The free motion of the internal orientation  
on $\mathbb{C}P^1 \simeq S^2$.}
\label{fig:s2}
\end{figure}

In the effective action of vortices, 
the charges of vortices are given by 
\beq
\vec Q = \frac{\p^2 K}{\p \beta^I \p \bar\phi{}^j} 
\vec \xi^{\,I} \dot{\bar\phi}{}^j
 + {\rm (c.c.)},
\eeq
where $\vec \xi^{\,I}$ are the holomorphic Killing vectors 
associated with $SU(2)$ rotation on the moduli space.
Explicit form of the $SU(2)$ triplet vector is given by
\beq
\vec \xi 
= \sum_{I=1}^k \vec \xi^{\,I} \frac{\p}{\p \beta_I},\quad
\vec \xi^I \equiv 
\left(-\frac{i}{2} (1 - \beta_I^2), \ 
\frac{1}{2} (1 + \beta_I^2), \ 
i \beta_I\right).
\label{eq:Killing}
\eeq
These charges have contributions from 
the free and interaction parts  
of the effective Lagrangian.
If all the vortices are isolated, 
contributions from the free part of the Lagrangian 
for each individual vortex are separately conserved.
In terms of $\vec n_I$, we define the charge of the $I$-th vortex by\footnote{
The normalization of $\vec Q_I$ is chosen so that 
they obey the half-integer quantization condition in the quantum theory.  
Although the coupling constant $g$ appears in the classical equations of motion
because of this normalization, it can be absorbed 
by rescaling $\vec Q_I$ appropriately. } 
\beq
\vec Q_I \equiv \frac{2\pi}{g^2} \vec n_I \times \dot{\vec n}_I.
\label{eq:Noether_charge}
\eeq
For the isolated vortices, 
the dynamics of the orientation is described by 
free equation of motion, 
which is nothing but the conservation law 
of the charge of each individual vortex 
\beq
\frac{d}{dt} \vec Q_I ~=~ \frac{2\pi}{g^2} \vec n_I \times \ddot{\vec n}_I ~=~ 0.
\eeq
This equation is equivalent to the geodesic equation on a sphere,
so that the trajectory of the orientation is a great circle, 
as shown in Fig.~\ref{fig:s2}: 
\beq
\vec n_I &=& \vec n_{I0} \, \cos ( \omega_I t ) 
+ \vec q_I \times \vec n_{I0} \, \sin ( \omega_I t ),
\label{eq:free_motion0}
\eeq
where 
$\vec n_{I0}$ and $\vec q_I$ are unit constant vectors satisfying 
$\vec n_{I0} \cdot \vec q_I = 0$. 
The charge of this solution is given by
\beq
\vec Q_I = \frac{2 \pi \omega_I}{g^2} \vec q_I.
\eeq
Note that we can always set $\omega_I \ge 0$ without loss of generality. 

\subsection{Features of the scattering of non-Abelian vortices
}\label{subsec:numerical}

As stressed in the previous subsection, 
the internal orientations and the conserved Noether charges 
make the dynamics of non-Abelian vortices 
quite different from that of the well-known ANO vortices. 
Since the internal orientations are continuous parameters, 
there exist continuously different initial conditions, 
which can make the dynamics quite complicated. 
In order to see how much 
the dynamics of non-Abelian 
vortices differs from that 
of the ANO vortices, 
we show several numerical solutions for the geodesic equations 
(\ref{eq:EOMposition}) and (\ref{eq:EOMorientation}).

The relative internal orientations between the first and 
second vortices are denoted as 
$\Delta \theta=\theta_1-\theta_2$, $\Delta \varphi=\varphi_1-\varphi_2$. 
We specify initial conditions at sufficiently past.  
We can always use the $SU(2)$ symmetry to set the initial values 
at sufficiently past as 
\beq
\varphi_{10} = - \varphi_{20}, \hs{10} 
\dot \varphi_{10} = \dot \varphi_{20} = 0.
\eeq
This initial condition corresponds to a pair of vortices 
whose orientations are rotating around great circles at the longitudes 
$\varphi_{10}$ and $\varphi_{20} = - \varphi_{10}$. 
Therefore, we can choose five initial conditions for the orientations: 
$\theta_I, \dot{\theta}_I~(I=1, 2)$ and $\Delta \varphi$. 

\paragraph{Scattering of vortices without initial $Q$-charges \\}
As the first example of the scattering of two non-Abelian vortices, 
we examine the case of $m_e=m_g$ and 
choose the following initial conditions 
for the relative orientation $\Delta \varphi_0$ and 
for the velocities of the orientations $\dot \theta_{I0}$:
\beq
\Delta \varphi_0 = 0, \hs{10} \dot{\theta}_{I0} = 0,~~~(I=1,2).
\eeq
In this setting, the vortices have no initial charges $\vec Q_{I0}=0$.
Because of these initial conditions, 
the variation of the internal orientation is very small 
before the vortices passes through the interaction region. 
\begin{figure}[h]
\begin{center}
\begin{tabular}{ccc}
\includegraphics[width=60mm]{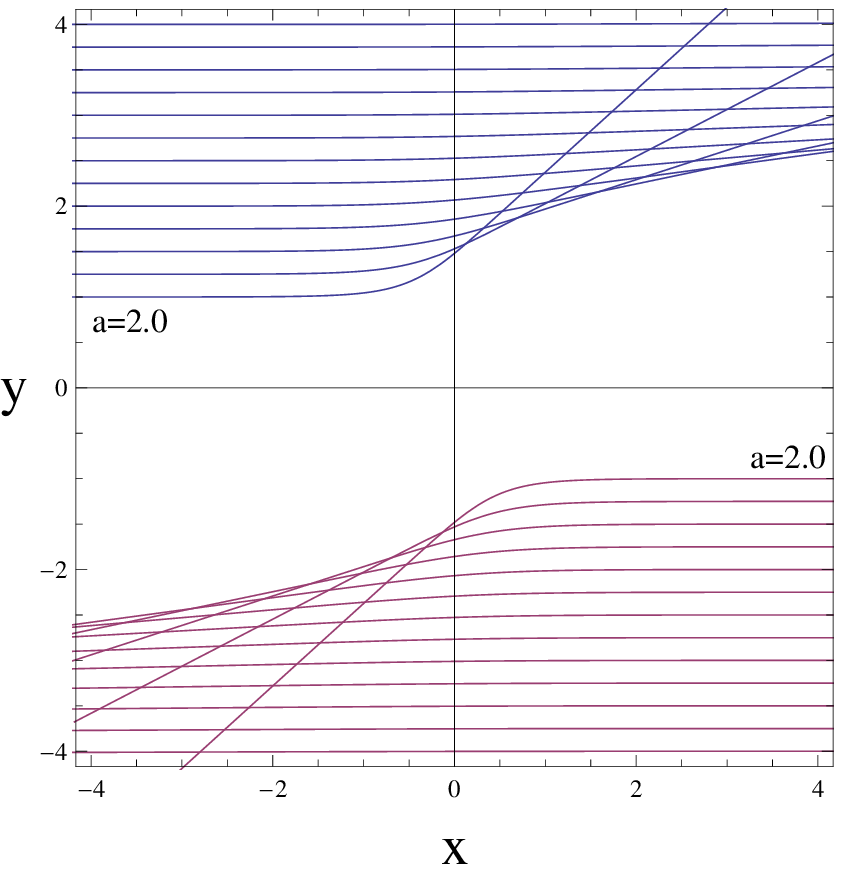} & \hs{10} 
&
\includegraphics[width=60mm]{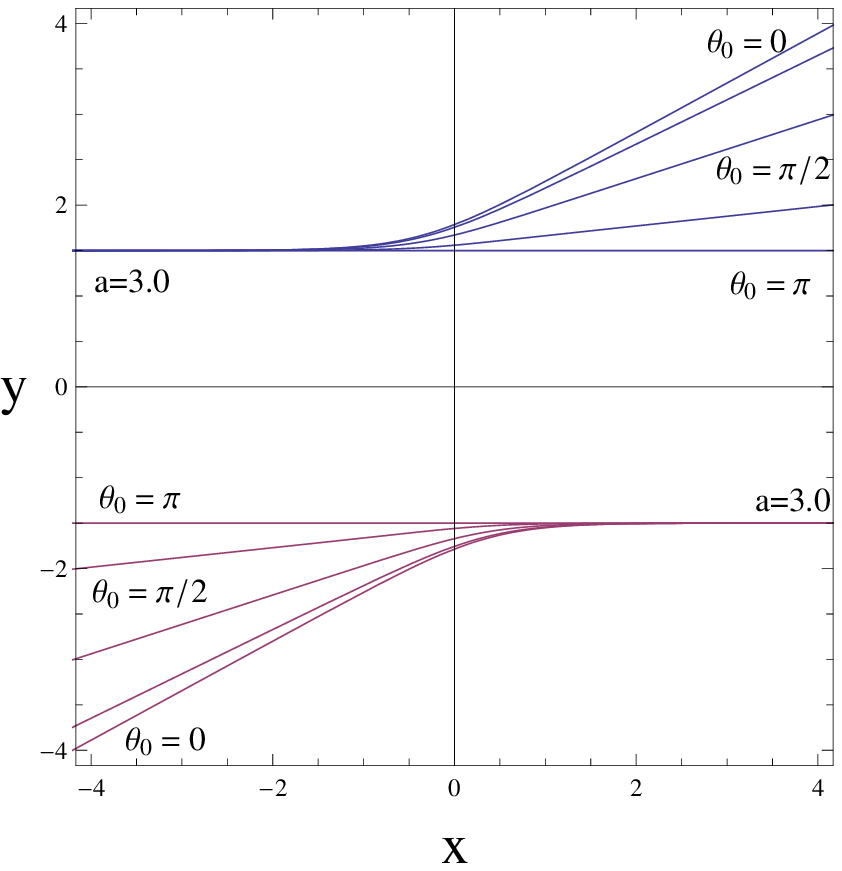} \\
(a) $a=2.0,~2.5, \cdots$,~~~$\Delta \theta_0 = \frac{\pi}{2}$ & \hs{10}
 &
(b) $a=3.0$,~~~$\Delta \theta_0=0,\frac{\pi}{4},\frac{\pi}{2},
\frac{3\pi}{4},\pi$
\end{tabular}
\caption{
Scattering orbits of non-Abelian vortices on the $z$-plane ($(x,y)$-plane)
in the case of $m_e=m_g=1$ ($c_e=c_g=1.708$),
for (a) various values of the impact parameter $a$
with the fixed initial relative orientation 
$\Delta\theta_0 =  \frac{\pi}{2}$, and 
(b) various values of the initial relative orientation $\Delta\theta_0$ 
with the fixed impact parameter $a=3.0$.
In the both cases the initial velocities of 
the orientations are zero ($\vec Q_I = 0$).
No interaction exists for $\Delta \theta_0=\pi$ (anti-parallel orientations). 
Otherwise, the interaction is always repulsive 
with reaching the maximum at $\Delta \theta_0=0$ (parallel orientations), 
where the interaction reduces to that of the ANO vortices.
}
\label{fig:test1}
\end{center}
\end{figure}

In Fig.~\ref{fig:test1}-(a),
we show the scattering orbits for the initial relative 
orientation $\Delta \theta_0=\pi/2$ by 
changing the impact parameter $a$ as 
$a = 2.0,\ 2.5,\ 3.0,\ \cdots$. 
One can clearly see that 
the moving vortices feel the repulsive geodesic force between each other.
At a glance, this result appears to be very similar to 
the scattering of the ANO vortices, 
in which case the scattering is repulsive and 
the orbits are uniquely determined 
if the initial velocity and the impact parameter are fixed.
However, it is actually different. 
See Fig.~\ref{fig:test1}-(b) 
where the impact parameter is fixed at $a=3$ 
and the initial relative orientation is varied as 
$\Delta \theta_0=0,\frac{\pi}{4},\frac{\pi}{2},\frac{3\pi}{4},\pi$. 
When $\Delta\theta_0= 0$ ($\Delta\theta_0 = \frac{\pi}{4},\frac{\pi}{2}...)$, 
they scatter (almost) in the same way as the Abelian vortices. 
On the other hand, non-Abelian vortices just pass through 
each other without feeling any interactions for $\Delta\theta_0 = \pi$ 
(a pair of vortices with anti-parallel orientations). 
This behavior is markedly different from the scattering of ANO vortices. 
We thus have found that the scattering angle is sensitive 
to the relative orientation. 

The dependence on the relative orientation can be 
roughly understood from the fact that 
the interaction part of the K\"ahler potential 
$K_{\rm int}$ given in Eq.\,(\ref{eq:Kahlerpotential}) is proportional to 
$1 + \Theta_{12}$ for $m_g = m_e~(c_g=c_e)$. 
Since the relative orientation is almost unchanged 
from the initial condition until the vortices leave the interaction region, 
it follows that the scattering angle is also proportional to $1 + \Theta_{12}$.
For example, the scattering angle is maximized (vanishes) 
for $\Delta \theta_0 = 0$, ($\Delta \theta = \pi$) at which 
$1 + \Theta_{12} = 2$, ($1 + \Theta_{12} = 0$). 
We emphasize again that, in contrast to the non-Abelian vortex 
scattering, the scattering of the ANO vortices 
is uniquely determined with the initial velocity and 
the impact parameter fixed.

In this example, we have taken the initial condition 
without initial $Q$-charges. 
The numerical calculations show that 
the vortices are charged $\vec Q_1 = - \vec Q_2 \not = 0$ 
after the scattering, even if orientational moduli is initially static. 
We will see this phenomenon via an analytic discussion 
in section \ref{subsec:large}.

\paragraph{Scattering of vortices with maximized non-Abelian effect \\}
\begin{figure}[h]
\begin{center}
\begin{tabular}{ccc}
\includegraphics[width=60mm]{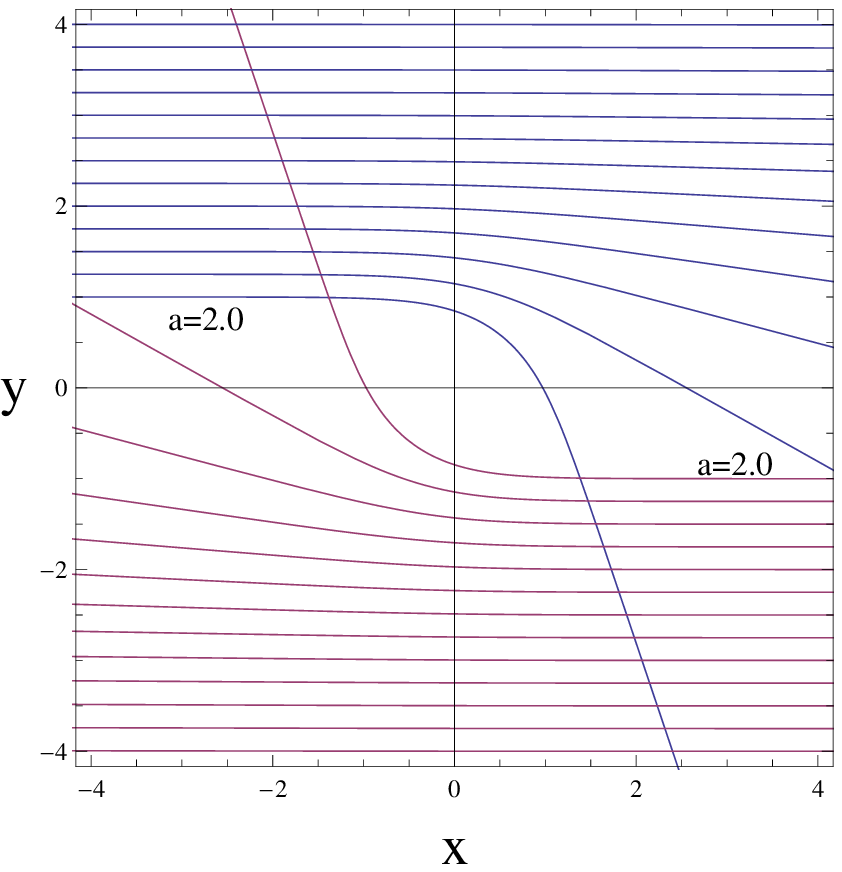} & \hs{10} &
\includegraphics[width=60mm]{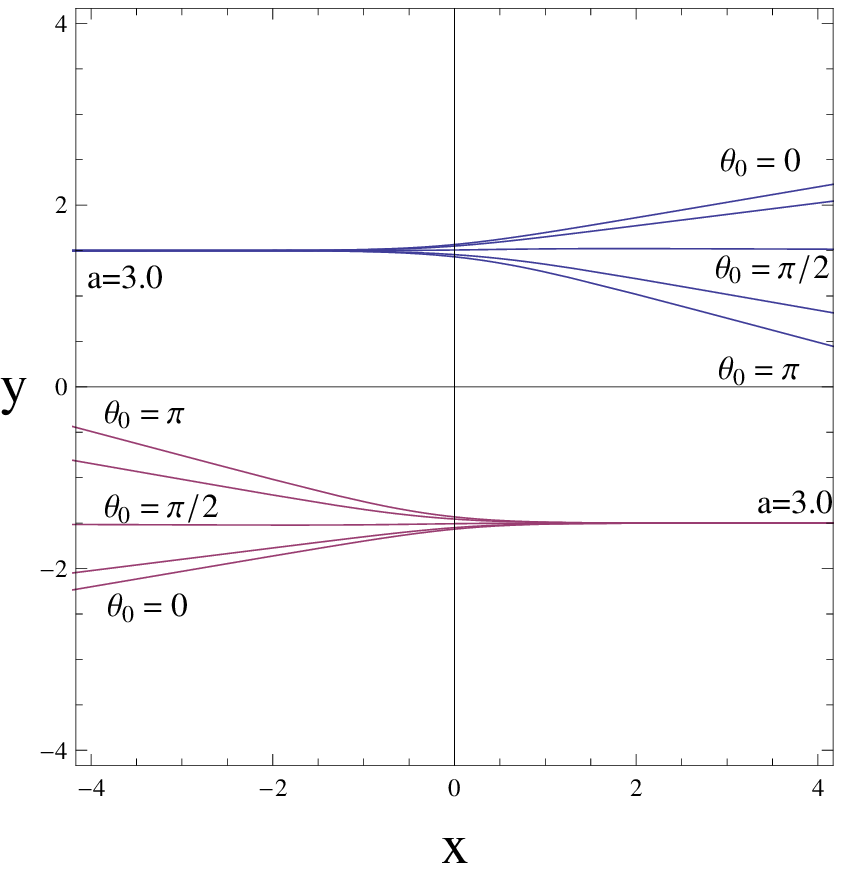} \\
(a) $a=2.0,~2.5, \cdots$,~~~$\Delta \theta_0 = \pi$ & \hs{10} &
(b) $a=3.0$,~~~$\Delta \theta_0=0,\frac{\pi}{4},\frac{\pi}{2},
\frac{3\pi}{4},\pi$
\end{tabular}
\caption{
Scattering orbits of non-Abelian vortices in the $z$-plane ($(x,y)$-plane)
for $m_e=\infty, m_g=1$ ($c_g=1.136$), with 
(a) various values of the impact parameter $a$ and 
the fixed initial relative orientation $\Delta\theta_0 = \pi$, 
and 
(b) various values of the initial relative orientation $\Delta \theta_0$ 
and the fixed impact parameter $a=3.0$.
In the both cases the other initial conditions are: 
$\Delta \varphi=0$ and $\dot\theta_I=0$.
}
\label{fig:test2}
\end{center}
\end{figure}
As the second example let us consider the case 
where non-Abelian effects are maximized. 
If the Abelian vector boson mass is sent to infinity\footnote{ 
Our original model reduces in the limit $m_e \to \infty$
to a ${\mathbb C}P^1$ nonlinear sigma model 
whose $SU(2)$ isometry is gauged. 
} 
$m_e \to \infty$, the Abelian part of the interaction 
K\"ahler potential $K_{\rm int}$, given by the first term in 
Eq.\,(\ref{eq:Kahlerpotential}), is highly suppressed and 
only the non-Abelian part, the second term, survives. 
The numerical results are shown in Fig.~\ref{fig:test2}.
Except for the Abelian mass $m_e$, the other parameters 
and the initial conditions are 
chosen to be the same as those for Fig.~\ref{fig:test1}.
Fig.~\ref{fig:test2}-(a) shows the 
scattering orbit for $\Delta \theta_0=\pi$ where 
the vortices clearly {\it attract} each other. 
This attractive force is 
a characteristic property of the non-Abelian case, 
which has not been seen in the Abelian case.
As before, the scattering angle is affected by 
the initial relative angle of the internal orientations, 
as shown in Fig.~\ref{fig:test2}-(b).
In the present case, the interaction 
is proportional to
$\Theta_{12} = \{1,0,-1\}$ for 
$\Delta\theta_0 = \{0,\frac{\pi}{2},\pi\}$, respectively.
Thus, the scattering angle changes its sign at 
$\Delta\theta_0 = \frac{\pi}{2}$. 
For $\Delta \theta_0 < \frac{\pi}{2}$, 
the vortices repel as in the Abelian case, 
while they attract for 
$\Delta \theta_0 > \frac{\pi}{2}$. 
The parallel orientation ($\Delta \theta_0=0$) gives the maximal repulsion 
and the anti-parallel orientation ($\Delta \theta_0=\pi$) 
gives the maximal attraction. 

\paragraph{Scattering of vortices with non-zero $Q$-charges \\}
\begin{figure}[h]
\begin{center}
\begin{tabular}{ccc}
\includegraphics[width=60mm]{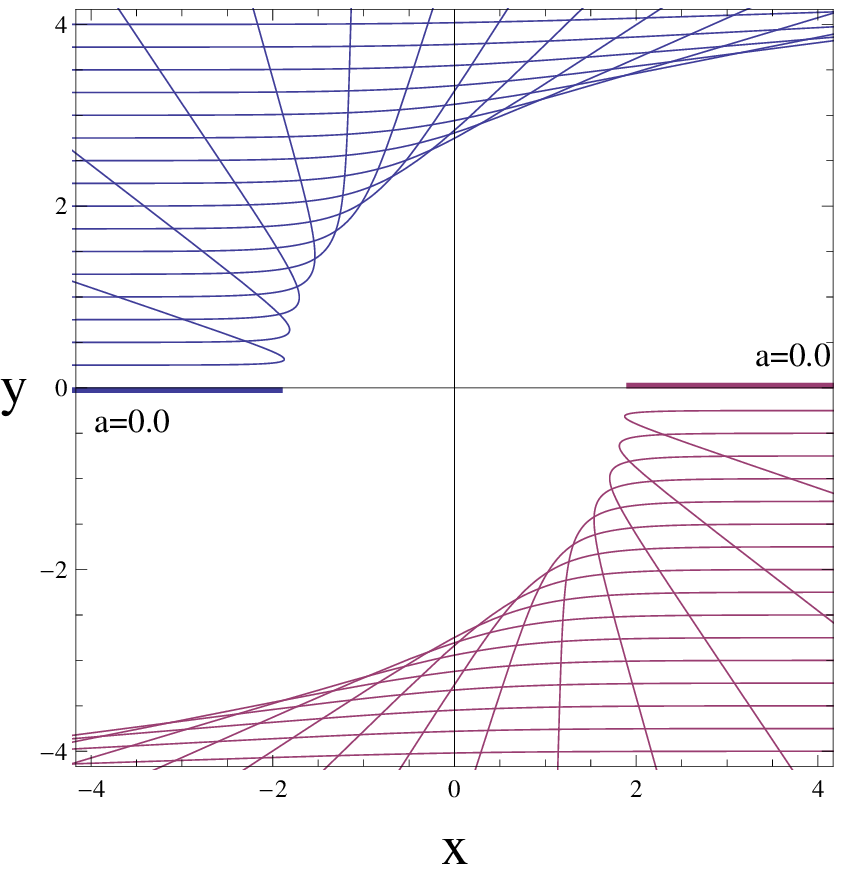} & \hs{10} &
\includegraphics[width=60mm]{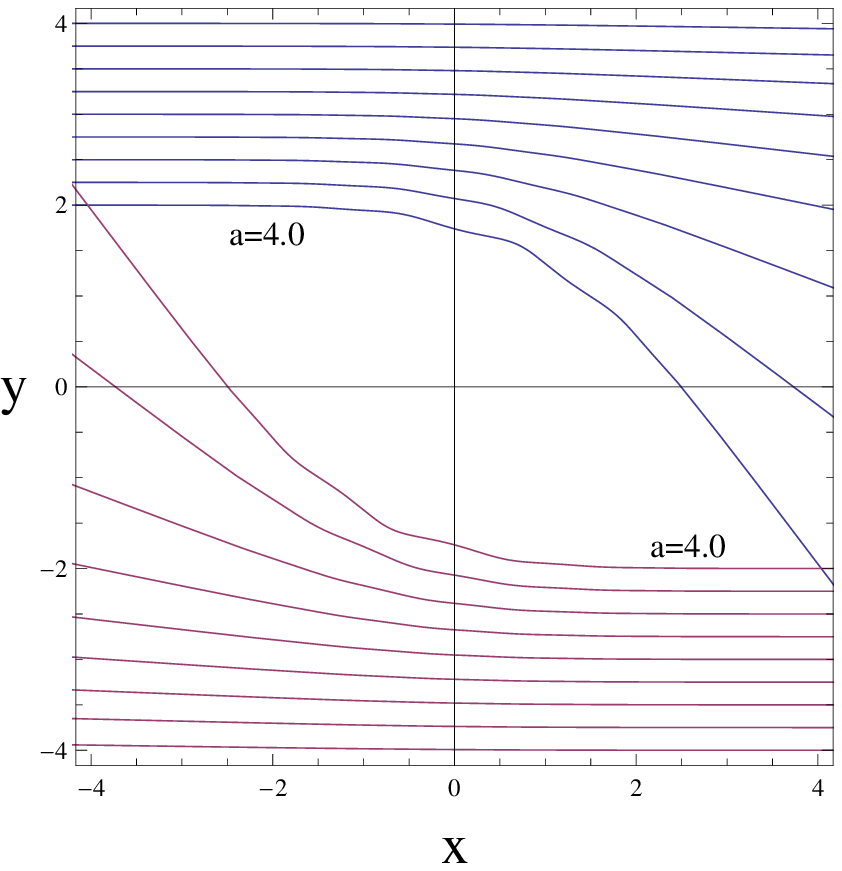} \\
(a) $a=0.0,~0.5,~\cdots$,~~~$\Delta \varphi_0 = \pi$ & \hs{10} &
(b) $a=4.0,~4.5,~\cdots$,~~~$\Delta \varphi_0=0$
\end{tabular}
\caption{
Scattering orbits of non-Abelian vortices with non-vanishing conserved 
charges for $\gamma_1 = - \gamma_2 = \frac{10}{3}$,
$m_e=m_g=1$ ($c_e=c_g=1.708$), 
with various values of the impact parameter $a$ and 
of the initial relative orientation $\Delta\varphi_0$. 
The other initial conditions are : 
$\theta_{10}=\theta_{20}=\frac{\pi}{2}$.
}
\label{fig:test3}
\end{center}
\end{figure}
The last interesting example is the scattering of non-Abelian 
vortices which have non-zero initial charges $\vec Q_I$ 
Since a repulsive (attractive) force 
works between the same (opposite) charges, 
the scattering is very different from that with vanishing 
charges. The results are shown in 
Fig.~\ref{fig:test3}, 
where $\gamma_I \equiv \dot \theta_{I0}/(m_gv)$ is defined by 
the initial relative velocity $v$ and 
the angular velocities $\dot \theta_{I0}$. 
Fig.\,\ref{fig:test3}-(a) shows the scattering orbits of 
vortices with the same initial $Q$-charges. 
Because of the repulsive force, the vortices recoil even for $a=0$.
On the other hand, the vortices with opposite charges feel attractive force
as shown in Fig.\,\ref{fig:test3}-(b). 
We can see that their orbits are slightly wavy. 
This is due to oscillations of the forces caused 
by the rotations of the orientations. 
In the next section, we will discuss an ``average" 
over the rapid motions of the orientations to find out 
effective forces between non-Abelian vortices with $Q$-charges.

\subsection{Forces between non-Abelian vortices}\label{subsec:force}

Now let us examine the geodesic forces between two vortices 
induced by motions of the moduli parameters. 
To this end, let us first rewrite the equations of motion 
Eqs.\,(\ref{eq:EOMposition}) and (\ref{eq:EOMorientation})  
in terms of $\vec n_I$ as\footnote{
Note that the equation of motion $\vec n \times \ddot{\vec n} = \vec A$
is equivalent to $\ddot {\vec n}= \vec A \times \vec n -\vec n |\dot {\vec n}|^2$ for vectors $\vec n$ and $\vec A$ such that
$|\vec n|^2 = 1$, $\vec n \cdot \vec A=0$.
}
\beq
\ddot z_{12} &=& - 
\left[ \frac{c_e^2}{m_e} K_1(m_e |z_{12}|) 
+ \frac{c_g^2}{m_g} (\vec n_1 \cdot \vec n_2) 
K_1(m_g|z_{12}|) \right] \frac{\bar z_{12}}{|z_{12}|} 
(m_g \dot z_{12})^2 \notag 
\\
&{}& 
+ \frac{2 c_g^2}{m_g} K_0(m_g|z_{12}|)
\left(\vec n_1\cdot \vec \alpha_2+\vec n_2\cdot \vec \alpha_1\right) 
m_g \dot z_{12}  - \frac{2c_g^2}{m_g} K_1(m_g |z_{12}|) 
\vec \alpha_1\cdot \vec \alpha_2
\frac{z_{12}}{|z_{12}|}, \phantom{\Bigg[} 
\label{eq:eomz12}
\eeq
\beq
\vec n_1 \times \ddot{\vec n}_1 
&=& {\rm Re} \bigg[ \ \, \frac{1}{2} c_g^2 K_2(m_g |z_{12}|) 
\ \vec n_1 
\times \left( \vec n_2 - i \vec n_1 \times \vec n_2 \right) 
\left( \frac{\bar z_{12}}{|z_{12}|} m_g \dot z_{12} \right)^2 
\notag \\
&{}& \phantom{\frac{1}{2}} + 2 c_g^2 K_1(m_g|z_{12}|)\, 
\, i \, \vec \alpha_1 
\left(\vec n_1 \cdot \vec n_2 
-\frac{{\vec \alpha}_1^\dagger\cdot \vec
\alpha_2}{|\vec \alpha_1|^2}\right)
\frac{\bar z_{12}}{|z_{12}|} m_g \dot z_{12} 
\phantom{\frac{1}{2}} \notag \\
&{}& \phantom{\frac{1}{2}} - \ \, c_g^2 K_0(m_g |z_{12}|) 
\,i\, \vec \alpha_1 \left(\vec n_2 \cdot \vec \alpha_1 
+ 2 \vec n_1 \cdot \vec \alpha_2 \right) \bigg]. 
\label{eq:eomq1} 
\eeq
where $\vec \alpha_I~(I=1,2)$ is a complex three-vector defined by
\begin{eqnarray}
\vec \alpha_I\equiv \dot{\vec n}_I-i\vec n_I\times \dot{\vec n}_I.
\end{eqnarray}
Note that the equation of motion for $\vec n_2$ can be obtained by
exchanging $\vec n_1$ and $\vec n_2$ in Eq.\,\eqref{eq:eomq1}.

The geodesic force given in the right-hand side of 
Eq.\,(\ref{eq:eomz12}) shows that the 
motion in orientational moduli space generally 
induces a force between the vortices 
even if there is no spatial motion initially. 
Similarly, Eq.\,(\ref{eq:eomq1}) shows that 
a spatial motion induces a motion in orientational moduli space, 
even if orientational moduli is initially static. 
This implies that the kinetic energy in spatial motion 
and in orientational moduli transmute each other.

\subsubsection{Average over rapid motion of internal orientations}
 
As we have seen in the numerical example Fig.\,\ref{fig:test3}-(b), 
the sign and magnitude of forces between them 
oscillate due to the rotations of $\vec n_I$.
Since motion in orientational moduli is (almost) periodic, 
its physical effect is best seen by averaging over the periods 
of two  orientational moduli individually.
Denoting this time average for 
the high-frequency modes
by $\langle \ \rangle $, 
we can express the averaged equation of motion for 
the relative position $z_{12}$ as
\beq
\langle \ddot z_{12} \rangle 
= \left<- \frac{z_{12}}{|z_{12}|} \left( c_e^2 m_e K_1(m_e |z_{12}|) 
\left( \frac{\dot z_{12} \bar z_{12}}{|z_{12}|} \right)^2 
- \frac{c_g^2 g^4}{2\pi^2 m_g} K_1(m_g |z_{12}|) \, 
\vec Q_1 \cdot \vec Q_2 \right) \right>. 
\label{eq:averagedEOMspatial}
\eeq
This approximation is valid only when the motion in space is slow 
compared to the velocities of the orientations,
\begin{eqnarray}
\omega_1,\omega_2\gg m_g|\dot z_{12}|\quad {\rm and~} 
\omega_1\not\approx \omega_2.\label{eq:largeomega}
\end{eqnarray} . 
The first term depends only on the velocities of the vortices 
just in the same way as the force between Abelian vortices. 
The second term is the dominant force induced by 
the motion of orientational moduli. 
Note that the order of the first term is $m_e|\dot z_{12}|^2$, 
while that of the second term is $\omega_1 \omega_2/m_g$.
Thus, in the case of $m_e \geq m_g$, 
the first term is negligible compared to 
the second term under the condition (\ref{eq:largeomega}). 
On the other hand, if $m_e < m_g$, 
the first term can be dominant
since $K_0(m_e|z_{12}|) \gg K_0(m_g|z_{12}|)$ asymptotically. 

It has been shown that the interaction of non-Abelian 
vortices are well described by regarding them as 
point-like sources of the Higgs fields and 
massive vector fields \cite{Fujimori:2010fk}. 
The motion of the orientation 
of the $I$-th vortex induces the following 
(color) electric charge distribution\footnote{
The charges $\vec Q_I$ themselves are not the electric charges 
but the charges of $SU(2)$ global symmetry. 
The total electric charges are always zero 
even for vortices with rotating orientations. }
and the massive vector field 
\beq
j_0^I = \frac{2g^2 c_g}{m_g^2} \ \vec Q_I \cdot 
\vec \sigma \ \p_z \p_{\bar z} \delta^2(z-z_I), \hs{10} 
W_0^I = \frac{g^2 c_g}{4\pi} \ \vec Q_I \cdot 
\vec \sigma \ K_0(m_g |z - z_I|).
\eeq
Therefore, the electrostatic potential between $I$-th 
and $J$-th vortices is given by
\beq
V_{IJ} = \int d^2 x \tr \left[ \frac{1}{g^2} j_0^I W_0^J \right] 
= \frac{c_g^2 g^2}{4\pi} \vec Q_I \cdot \vec Q_J \ K_0(m_g |z_{IJ}| ).
\label{eq:QQ}
\eeq
The second term in the averaged equation of motion 
Eq.\,\eqref{eq:averagedEOMspatial} can be attributed to 
this ``electrostatic interaction".
This force is repulsive (attractive) 
if the inner product of the charges of two vortices is positive (negative). 
Namely the vortices with aligned charges repel each other, 
whereas those with disaligned charges attract each other. 
These phenomena can be seen in the numerical calculations 
in Fig.\,\ref{fig:test3}.

\begin{figure}
\begin{center}
\includegraphics[height=4.5cm]{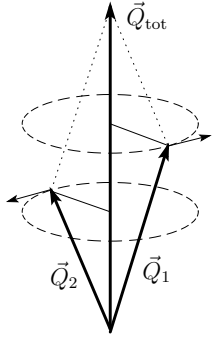}
\end{center}
\caption{An example of motions of the charge vectors 
$\vec Q_I$ for $\omega_I \gg m_g|\dot z_{12}|$.}
\label{fig:saisa}
\end{figure}

Let us next investigate the averaged equations of motion 
for the orientations. 
By using the charge vectors $\vec Q_I$ 
defined in Eq.\,\eqref{eq:Noether_charge}, 
the equation of motion of $\vec n_1$ Eq.\,(\ref{eq:eomq1}) reduces to 
\begin{eqnarray}
\left<\frac{d \vec Q_1}{dt}\right>&=& 
\frac{g^2c_g^2 }{2\pi} \left<K_0(m_g |z_{12}|)\, 
(\vec Q_2\times \vec Q_1)\right>,
\label{eq:averagedEOMQ1} 
\end{eqnarray}
where we have used 
$\left< \vec n_I \otimes \vec n_I^{\rm T} \right>
=\frac12\left< {\bf 1}-\vec q_I\otimes \vec q_I^{\rm T} \right>$. 
The equation for $\vec Q_2$ can be obtained just by exchanging 
the two vortices $(1\leftrightarrow 2)$. 
These equations describe precessions of the orientations, that is, 
motions along great circles with slowly moving axes. 
We can see from Eq.\,\eqref{eq:averagedEOMQ1} that the sum 
$\vec Q_{\rm tot}\equiv \left<\vec Q_1+\vec Q_2\right>$ 
and the inner product $\left<\vec Q_1\cdot \vec Q_2\right>$ are conserved.
By solving averaged equation of motion,
we find that each charge $\vec Q_I$ slowly rotates 
around $\vec Q_{\rm tot}$ as (see Fig.\,\ref{fig:saisa})
\begin{eqnarray}
\left<\vec Q_1\right> &\simeq& a \, \vec Q_{\rm tot} + \vec Q_{\rm osc} \cos \theta(t) + \frac{\vec Q_{\rm tot} \times \vec Q_{\rm osc}}
{|\vec Q_{\rm tot}|} \sin \theta(t), \\
\left< \vec Q_2 \right> &\simeq& b \, \vec Q_{\rm tot} -\vec Q_{\rm osc} \cos \theta(t) - \frac{\vec Q_{\rm tot}\times \vec Q_{\rm osc} }
{|\vec Q_{\rm tot}|} \sin \theta(t),
\end{eqnarray}
where $a + b = 1$, $\vec Q_{\rm osc} \cdot \vec Q_{\rm tot}=0$ 
and
\begin{eqnarray}
\theta(t) &=&\frac{g^2c_g^2 }{2\pi} |\vec Q_{\rm tot}| \int^t dt \ 
K_0(m_g |z_{12}|).
\end{eqnarray}
We can see that the angular frequency of the rotation 
is quite small $\dot \theta \ll \omega_{1,2}$.

In contrast to the Abelian vortex, 
the non-Abelian vortex has the characteristic feature 
that the interactions depend on the relative internal orientations. 
Since the inner product $\vec Q_1\cdot\vec Q_2$ is 
almost conserved during the scattering process, the 
interaction can be, for instance, kept to be attractive 
by choosing near anti-parallel orientations. 
In such a situation, one can imagine a possibility for two 
non-Abelian vortices to be bound together. 
However, it is not the case. It is because the balanced point 
for the exponentially decaying attractive force ($\sim e^{m_g |z_{12}|}$)
with the centrifugal force ($\sim 1/|z_{12}|^2$) is unstable. 
Of course, this conclusion is valid only for the 
well-separated vortices at distances $|z_{12}|>1/m_g$. 
Although one may expect that bound states can exist when $|z_{12}|<1/m_g$,
such a region is out of range of validity of our approximation.  
In Sec.~\ref{sc:dyonicVortex}, we will see that 
a bound state of two (dyonic) vortices exists 
in a mass deformed theory where a certain extra force, 
which is not a geodesic force, is induced by a mass term.

\subsubsection{Limit of slowly moving internal orientations}
 
Let us next consider the opposite limit of Eq.\,(\ref{eq:largeomega}), 
\beq
\omega_I \ll m_g|\dot z_{12}|.
\eeq
Now the motion of the internal orientations is very slow 
compared to the spatial motion, so that we can safely  
neglect $\dot{\vec{n}}_I$ in the equations. 
Thus the first lines of Eq.\,(\ref{eq:eomz12}) and 
Eq.\,(\ref{eq:eomq1}) give dominant contributions. 
Depending on the mass scales $m_g$ and $m_e$, 
the dominant contributions in Eq.\,(\ref{eq:eomz12}) are given by
\begin{eqnarray}
\ddot z_{12} \hs{-2} &\approx& \hs{-2}\left\{ 
\begin{array}{lcl}
\displaystyle- \frac{c_e^2}{m_e} K_1(m_e |z_{12}|)
\frac{\bar z_{12}}{|z_{12}|} (m_e \dot z_{12})^2	
&\qquad& 
{\rm for~}\quad m_e<m_g \\
\displaystyle - \frac{c_e^2}{m_e}
\left(1+\vec n_1 \cdot \vec n_2\right) K_1(m_e |z_{12}|)
\frac{\bar z_{12}}{|z_{12}|} (m_e \dot z_{12})^2		
&\qquad& 
{\rm for~} \quad m_e=m_g \\
\displaystyle - \frac{c_g^2}{m_g} (\vec n_1 \cdot \vec n_2) 
K_1(m_g|z_{12}|) \frac{\bar z_{12}}{|z_{12}|} (m_g \dot z_{12})^2
&\qquad& {\rm for~}\quad m_e>m_g
\end{array}\right.   
\end{eqnarray}
Since the interaction for $m_e < m_g$ is equivalent 
to that of the Abelian case, the vortices receive repulsive force 
in the scattering process. 
For $m_e > m_g$, the forces depend on $\vec n_1\cdot\vec n_2$. 
When $\vec n_1\cdot\vec n_2 > 0 $ ($\vec n_1\cdot\vec n_2 < 0 $), 
they repel (attract) each other and
the interaction accidentally vanishes at $\vec n_1 \cdot\vec n_2 = 0$. 
Although the interaction depends on $\vec n_1 \cdot \vec n_2$ 
also in the case of $m_e = m_g$, 
it is qualitatively similar to the Abelian case, 
that is,  
the vortices receive the repulsive force 
except for the case $\vec n_1 \cdot \vec n_2 = -1$. 
This explains the behaviors of the scattering given in 
Figs.~\ref{fig:test1} and \ref{fig:test2}.

\subsubsection{Internal orientations at a head-on collision}
As a special case, let us consider a pair of vortices 
which are going to collide head-on with each other
\beq
\dot z_{12} = - v \frac{z_{12}}{|z_{12}|}, \hs{10} \dot{\vec n}_1 ,
\dot{\vec n}_2 \simeq 0,\quad v > 0.
\eeq
From the equation of motion for the orientation 
Eq.\,\eqref{eq:eomq1}, we find that 
\beq
\vec n_1 \times \ddot{\vec n}_1 \simeq 
c_g^2 m_g^2 v^2 K_2(m_g |z_{12}|) \ \vec n_1 \times \vec n_2.
\label{eq:collision}
\eeq
This equation implies that the orientations of the vortices tend to align 
before a head-on collision. This result for well-separated
vortices naturally extends a similar result obtained by an analysis 
around a vicinity of coincident vortices \cite{Eto:2006db}.

\subsection{Large impact parameter}\label{subsec:large}

If two vortices are completely separated, 
interactions between them can be neglected, 
and the solution of the equations of motion is given by the free motion 
(moving parallel to the $x$-axes)
\beq
z_{12} = v t + i a, \hs{10} 
\vec n_I&=&\vec c_I e^{i\omega_I t}+\vec c_I^{\ \ast} e^{-i\omega_I t}.
\label{eq:free_motion}
\eeq
Here we have rewritten the solution Eq.\,(\ref{eq:free_motion0}) 
in terms of a complex vector $\vec c_I$ which 
is related to $\vec n_{I0}$ and $\vec q_I$ as
\begin{eqnarray}
\vec c_I = \frac{1}{2} \left(\vec n_{I0}-i\vec q_I\times \vec n_{I0}\right)
\quad \leftrightarrow \quad 
\vec n_{I0} = \vec c_I + \vec c_I^{\ \ast}, \quad 
\vec q_I = -2i \vec c_I \times \vec c_I^{\ \ast}.
\end{eqnarray}
Let us here consider the case of large impact parameter $a$.
If two vortices are far apart, 
their interactions are exponentially suppressed and 
the deviation from the free motion is small during the scattering process. 
Therefore we can safely evaluate their interactions by 
approximating the right-hand side of the equations of motion 
in Eqs.\,(\ref{eq:eomz12}) and (\ref{eq:eomq1}) 
by inserting the free motion in Eq.\,(\ref{eq:free_motion}).  

\subsubsection{Scattering angle}
Substituting the free motion Eq.\,(\ref{eq:free_motion}) 
into the right-hand side of Eq.\,\eqref{eq:eomz12}, 
we obtain the approximated equation of motion 
for the relative position $z_{12}$ 
in which all the forces are known functions of time. 
Then, we can evaluate the total change in the relative velocity 
by integrating the forces from $t=-\infty$ to $t=\infty$
\beq
\Delta \dot z_{12} = \int_{-\infty}^{\infty} \ddot z_{12} \, dt. 
\eeq
This can be carried out explicitly by using 
the following formulas of Fourier transformations
\beq
\int_{-\infty}^\infty dt \, K_0(m \sqrt{(vt)^2+a^2}) e^{i \omega t} \hs{-2} &=& \hs{-2} \frac{\pi}{\sqrt{(mv)^2+\omega^2}} e^{- \frac{a}{v} \sqrt{(mv)^2+\omega^2}}, \\
\int_{-\infty}^\infty dt \frac{vt \pm i a}{\sqrt{(vt)^2+a^2}} K_1(m \sqrt{(vt)^2+a^2}) e^{i \omega t} \hs{-2} &=& \hs{-2} \frac{\pi i}{mv} \left[ \frac{\omega}{\sqrt{(m v)^2 + \omega^2}} \pm 1 \right] e^{- \frac{a}{v} \sqrt{(mv)^2+\omega^2}}. 
\eeq
Eventually, we obtain the following deviation of 
the relative velocity
\beq
\frac{\Delta \dot z_{12}}{2\pi i v} &=& 
 \frac12 c_e^2 e^{- m_e a} +  c_g^2 \gamma_1 \gamma_2 (\vec q_1 \cdot \vec q_2) e^{-m_g a} \phantom{\Big[} \label{eq:dz12} \\
&{}& - c_g^2 \gamma_2 \left( \sqrt{1 + \gamma_1^2} \, \vec q_2 + i \gamma_1 \, \vec q_1 \times \vec q_2 \right) \cdot \vec n_{10} \, e^{-m_g a \sqrt{1 + \gamma_1^2}} \notag \\
&{}& -  c_g^2 \gamma_1 \left( \sqrt{1 + \gamma_2^2} \, \vec q_1 - i \gamma_2 \, \vec q_1 \times \vec q_2 \right) \cdot \vec n_{20} \, e^{-m_g a \sqrt{1 + \gamma_2^2}} \notag \\
&{}& +  c_g^2 (1 + 2 \gamma_1 \gamma_2) \left( {\rm Re}( \vec c_1 \cdot \vec c_2 
) + \frac{\gamma_1+\gamma_2}{\sqrt{1 + (\gamma_1 + \gamma_2)^2}} i \, {\rm Im}( \vec c_1 \cdot \vec c_2 ) \right) e^{- m _g a \sqrt{1 + (\gamma_1 + \gamma_2)^2}} \notag \\
&{}& +  c_g^2 (1 - 2 \gamma_1 \gamma_2) \left( {\rm Re}( \vec c_1 \cdot \vec c_2^{\ \ast} ) + \frac{\gamma_1-\gamma_2}{\sqrt{1 + (\gamma_1 - \gamma_2)^2}} i \, {\rm Im}( \vec c_1 \cdot \vec c_2^{\ \ast} ) \right) e^{- m _g a \sqrt{1 + (\gamma_1 - \gamma_2)^2}}. \notag 
\label{eq:z12}
\eeq 
Note that the frequencies $\omega_I$ appear only through the following ratio 
\begin{eqnarray}
\gamma_I \equiv \frac{\omega_I}{m_g v}.
\end{eqnarray}
This follows from the invariance of the ratio $\Delta \dot z_{12}/v$ 
under the time rescaling $t\to \lambda t$, 
which arises from the property of the geodesic equation 
(\ref{eq:geodesic_eq}) itself. 
This ratio $\gamma_I$ measures how much the vectors 
$\vec n_I$ rotate in the typical time scale of the interaction 
$\Delta t \approx 1/m_g v$. 
For generic values of the ratios $\gamma_I$ and the vectors $\vec q_I$, 
only the first line in Eq.\,(\ref{eq:dz12}) is dominant,
\beq
\Delta \dot z_{12} \approx 
\left\{ \begin{array}{ccc} 
\pi i v c_e^2 e^{- m_e a} & \qquad & m_e < m_g \\ 
2 \pi i v c_g^2 \gamma_1 \gamma_2 (\vec q_1 \cdot \vec q_2) 
e^{-m_g a} & \qquad & m_g < m_e \end{array} \right..
\label{eq:nonresonant}
\eeq
In the case of $m_e < m_g$, the leading contribution comes from 
the first term in the averaged equation of motion 
Eq.\,\eqref{eq:averagedEOMspatial},
while the second term is dominant for $m_g < m_e$.

The free-motion approximation is valid for orbits 
with small scattering angles. 
This means that the kinetic energy of the spatial motion 
should be sufficiently larger than 
that of the internal orientations.
If vortices with non-zero charges have small relative velocities, 
the interaction Eq.\,\eqref{eq:QQ} lasts 
for a long time interval $\Delta t \approx 1/(m_g v)$ 
and the deviation of the trajectory 
from the free motion becomes large. 
Therefore $\gamma_1$ and $\gamma_2$ should be 
generically small so that 
\beq
\gamma_1 \gamma_2 \, e^{- m_g a} ~\ll~ 1.
\eeq  
\begin{figure}[h]
\begin{center}
\includegraphics[width=60mm]{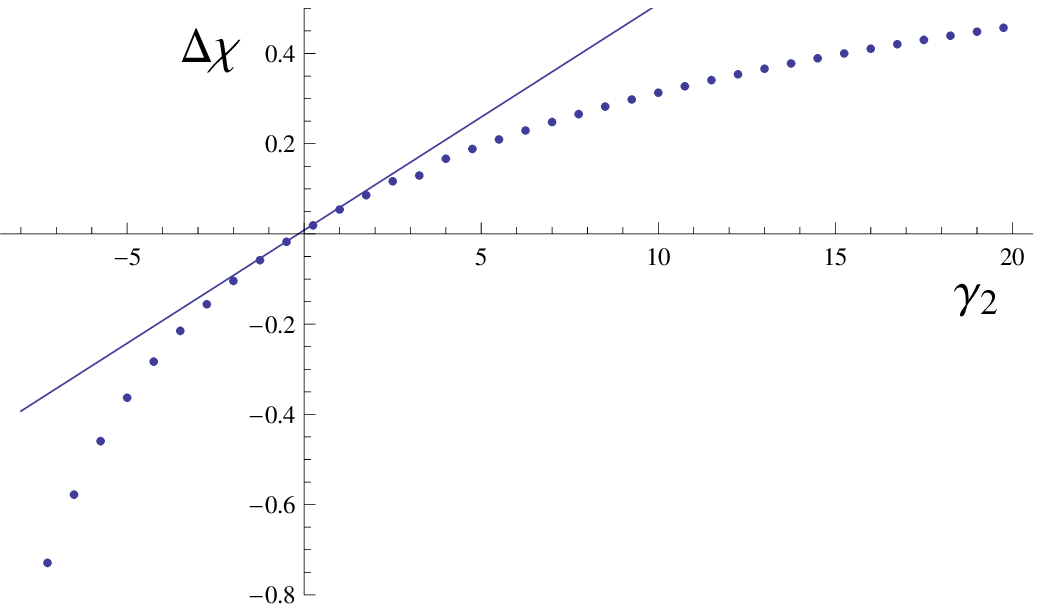}\qquad \qquad
\includegraphics[width=50mm]{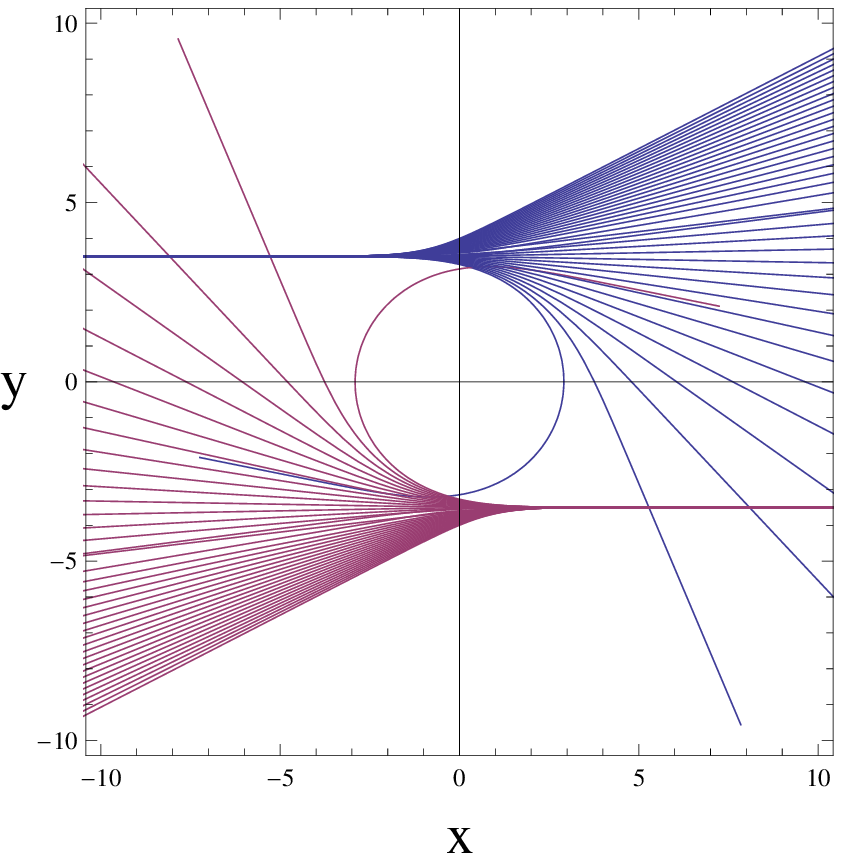}\\
\hs{20} (a) $\Delta \chi$\hs{50}(b) Orbits in the $z$-plane \vspace{-0.5cm}
\end{center}
\caption{\small (a) 
Scattering angles $\Delta \chi$ as a function of $\gamma_2$, 
and (b) the corresponding orbits of the vortices in the $z$-plane 
(($x,y$)-plane) for $m_e=m_g=1$, $a=7$, $\vec q_1 = \vec q_2 = (1,0,0)$, 
$\vec n_{10}=(0,1,0)$, $\vec n_{20}=(0,0,1)$ and
$\gamma_1=3$. 
}  
\label{fig:angles}
\end{figure}
In Fig.~\ref{fig:angles}-(a), 
we compare the scattering angles obtained by numerical calculations 
and those with the free-motion approximation.
One finds that the free-motion approximation is indeed 
valid only for the scattering with small scattering angles. 
We also show the numerical results of 
the scattering with various initial conditions 
in Fig.~\ref{fig:angles}-(b). 
Among them, there are orbits whose scattering angles exceed $\pi/2$. 
For such collisions, the free-motion approximation cannot be applied.

The terms with $\gamma_I$ in the exponents 
(terms other than the first line) in Eq.\,(\ref{eq:dz12}) 
are contributions from the oscillating forces and 
become smaller for larger values of $\gamma_I$. 
Although those forces give the subdominant contributions,  
some of the subleading terms in Eq.\,\eqref{eq:dz12} 
show resonant behaviors and become comparable to the leading term 
at $\omega_1 = 0$, $\omega_2=0$ or $\omega_1 = \omega_2$. 
This is because the corresponding forces in the equation of motion
Eq.\,\eqref{eq:eomz12} do not oscillate for these values of frequencies, 
that is, the forces are not averaged in the case of vortices 
with synchronized orientations. 
\begin{figure}[h]
\begin{center}
\includegraphics[width=50mm]{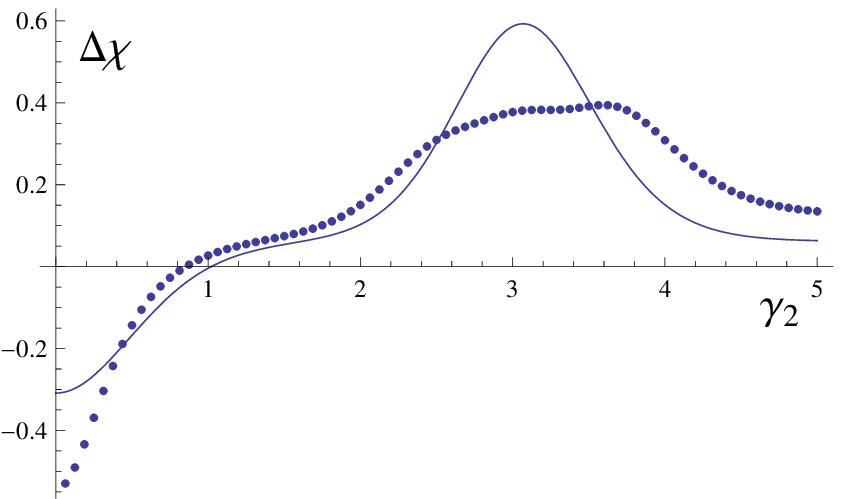}\qquad 
\includegraphics[width=50mm]{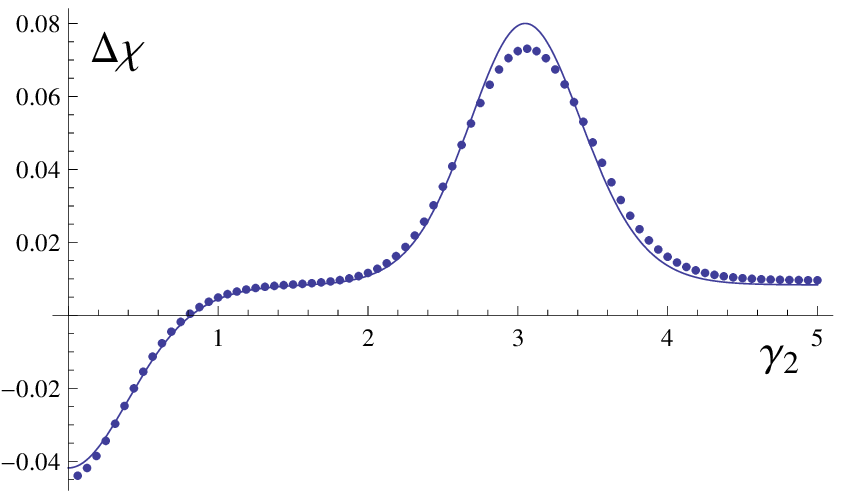}\qquad 
\includegraphics[width=50mm]{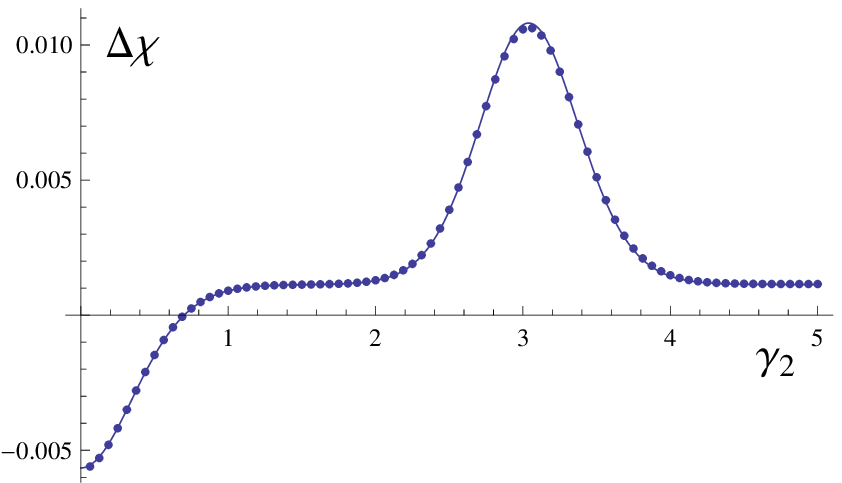}\\
(a) $a=5$\hs{40} (b) $a=7$ \hs{35} (c) $a=9$ \vspace{-0.5cm}
\end{center}
\caption{\small 
Scattering angles as a function of $\gamma_2$ 
for $m_g=m_e=1$, $a=5,7,9$, $\gamma_1=3$, $\vec q_1=\vec n_{20}=(1,0,0)$ 
and $\vec q_2=\vec n_{10}=(0,1,0)$. 
The numerical results and those of free-motion approximation 
are denoted by dotted lines and solid lines, respectively. 
The resonances can be seen at $\gamma_2=0$ and $\gamma_2=\gamma_1=3$.}  
\label{fig:resonances}
\end{figure}
In the case of $m_e > m_g$ with $\vec q_1\cdot \vec q_2=0$, 
the leading term Eq.\,(\ref{eq:nonresonant}) vanishes
and the terms with the resonant behavior become dominant 
as shown in Fig.~\ref{fig:resonances}. 

\subsubsection{Exchange of $Q$-charges}

In the scattering process, the 
charge of individual vortex 
in Eq.\,(\ref{eq:Noether_charge}) is not conserved and 
its time dependence is given by the equation of motion 
for the orientation.  
Substituting the free motion Eq.\,\eqref{eq:free_motion} 
into the equation of motion Eq.\,\eqref{eq:eomq1} 
and integrating the right-hand side, we obtain 
the increment of the charge of the first vortex 
$\Delta \vec Q_1~(=- \Delta \vec Q_2)$ as
\beq
\Delta \vec Q_1 &=& - \frac{2 \pi^2 c_g^2 m_g v}{g^2} \left( \gamma_1
\gamma_2 \ \vec q_1 \times \vec q_2 \ e^{- m_g a} + \gamma_2 \ \vec q_2
\times \vec n_{10} \ e^{- m_g a \sqrt{1+\gamma_1^2}} - \gamma_1 \ \vec
q_1 \times \vec n_{20} \ e^{- m_g a \sqrt{1+\gamma_2^2}} \right) \notag
\\
&{}& - \frac{2\pi^2 c_g^2 m_g v}{g^2} \frac{1 + 2 \gamma_1
\gamma_2}{\sqrt{1+(\gamma_1+\gamma_2)^2}} 
{\rm Re} [\vec c_1\times \vec c_2]
e^{- m_g a \sqrt{1+(\gamma_1+\gamma_2)^2}} \notag \\
&{}& - \frac{2\pi^2 c_g^2 m_g v}{g^2} \frac{1 - 2 \gamma_1
\gamma_2}{\sqrt{1+(\gamma_1-\gamma_2)^2}} 
{\rm Re}[\vec c_1\times \vec c_2^{\ \ast}]
e^{- m_g a \sqrt{1+(\gamma_1-\gamma_2)^2}}. 
\label{eq:Q1}
\eeq
Here, we have used the following integration formula
\beq
\int_{-\infty}^\infty dt \bigg( 
\frac{vt \pm i a}{\sqrt{(vt)^2+a^2}} \bigg)^2 
K_2(m \sqrt{(vt)^2+a^2}) e^{i \omega t} \hs{40} \notag \\
\hs{40} = \frac{2\pi}{(mv)^2} 
\left( \mp \omega - 
\frac{\frac{1}{2} (mv)^2 + \omega^2}{\sqrt{(mv)^2+\omega^2}} 
\right) e^{- \frac{a}{v} \sqrt{(mv)^2 + \omega^2}}.
\eeq
For generic values of the angular velocities of the orientations $\gamma_I$, 
the leading contribution is given by
\beq
\Delta \vec Q_1 \approx - \frac{2 \pi^2 c_g^2 m_g v}{g^2} \ 
\gamma_1 \gamma_2 \ \vec q_1 \times \vec q_2 \ e^{- m_g a}.
\eeq
This contribution comes from the electric coupling Eq.\,\eqref{eq:QQ}
as in the case of $\Delta \dot z_{12}$. 
The resonant behavior of the subleading terms 
can also be seen at $\gamma_1 = 0$, $\gamma_2=0$ and 
$\gamma_1 = \pm \gamma_2$. 
In particular, if the vortices are initially uncharged 
($\gamma_1 = \gamma_2 = 0$), the increment $\Delta \vec Q_1$ is given by
\beq
\Delta \vec Q_1 = - \frac{\pi^2 c_g^2 m_g v}{g^2} \ 
\vec n_{10} \times \vec n_{20} \ e^{-m_g a}.
\eeq
Therefore, the vortices are charged 
after the scattering even if $\vec Q_I = 0$ initially.

\subsubsection{Exchange of energy} 
Let us now show that the energy can be transfered 
between two non-Abelian vortices through the scattering. 
This is another characteristic feature of the non-Abelian vortices. 
Since two Abelian vortices are identical objects 
without any internal degree of freedom, 
the energy transfer never occurs in the scattering of 
two Abelian vortices since only the elastic scattering 
is possible for two identical vortices. 

\begin{figure}[h]
\begin{center}
\includegraphics[width=60mm]{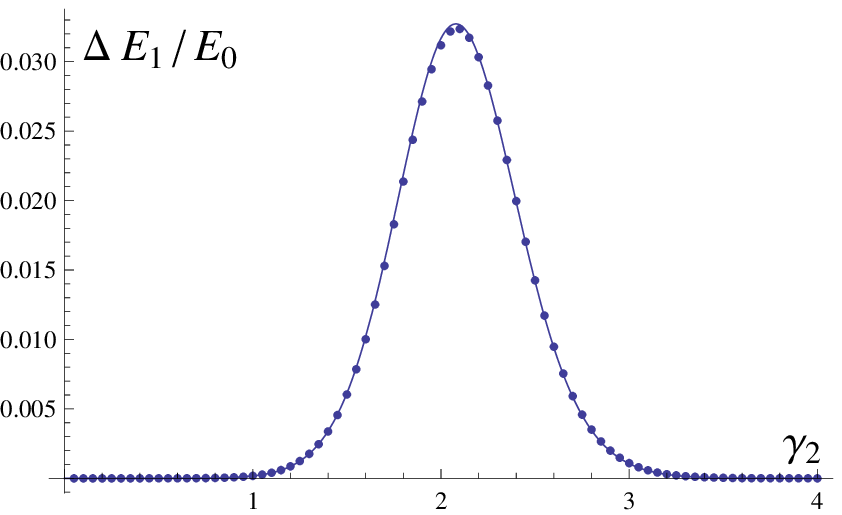}\qquad \qquad
\includegraphics[width=60mm]{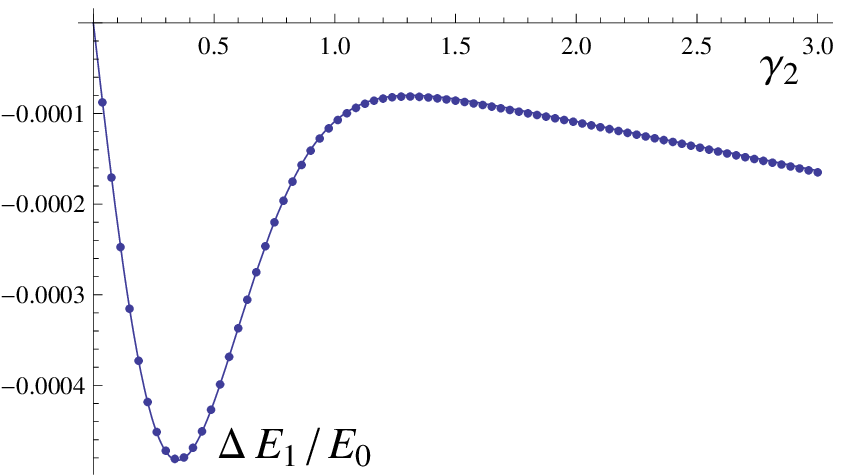}\\
(a) Resonance at $\gamma_1=\gamma_2$ \hs{30} (b) Resonance at $\gamma_2=0$
\end{center}
\caption{\small 
The energy transfer $\Delta E_1/E_0$ as a 
function of $\gamma_2$ ($m_g=m_e=1$). 
(a) $a=9$, $\gamma_1=2$, 
$\vec q_1 = \vec q_2=(1,0,0)$, $\vec n_{10}=(0,1,0)$ 
and $\vec n_{20}=(0,0,1)$, 
(b) $a=9$, $\gamma_1=1$, 
$\vec n_{10} = \vec n_{20}=(0,0,1)$, 
$\vec q_1 = (1,0,0)$ and $\vec q_2=(0,1,0)$. 
The resonances can be seen 
at $\gamma_2=\gamma_1$ in (a) and at $\gamma_2=0$ in (b). }  
\label{fig:EnergyResonance}
\end{figure}

The energy of an isolated vortex is given by 
the sum of the kinetic energies of the spatial motion and 
the orientation
\beq
E_I = \frac{\pi m_g^2}{g^2} |\dot z_I|^2 
+ \frac{g^2}{4\pi} |\vec Q_I|^2.
\eeq
Assuming that $\Delta \dot z_{12}$ and 
$\Delta \vec Q_1$ are sufficiently small, 
we can calculate the total change in energy, 
$\Delta E_1$ $(=-\Delta E_2)$, as 
\beq
\Delta E_1 &\approx& \phantom{+} \frac{\pi m_g^2}{g^2} 
(\Delta \dot z_1 \dot{\bar z}_1 + \dot z_1 \Delta \dot{\bar z}_1) 
 + \frac{g^2}{2\pi} \Delta \vec Q_1 \cdot \vec Q_1 \\
&\approx& -2\pi E_0\, c_g^2 \gamma_1 \gamma_2 \, 
(\vec q_1 \times \vec q_2 ) \cdot 
\left( \vec n_{10} \, e^{-m_g a \sqrt{1 + \gamma_1^2}} 
+ \vec n_{20} \, e^{-m_g a \sqrt{1 + \gamma_2^2}} \right) \notag \\
&{}& - 2\pi E_0\, c_g^2  \frac{(\gamma_1-\gamma_2)(1 +
2 \gamma_1 \gamma_2)}{\sqrt{1 + (\gamma_1 + \gamma_2)^2}} \, 
{\rm Im}[\vec c_1\cdot \vec c_2] \ e^{- m _g a
\sqrt{1 + (\gamma_1 + \gamma_2)^2}} \notag \\
&{}& + 2\pi E_0\, c_g^2  \frac{(\gamma_1+\gamma_2)(1 - 2
\gamma_1 \gamma_2)}{\sqrt{1 + (\gamma_1 - \gamma_2)^2}} \, 
{\rm Im}[\vec c_1\cdot \vec c_2^{\ \ast}] \ e^{- m _g a \sqrt{1
+ (\gamma_1 - \gamma_2)^2}}, \notag 
\eeq
where $E_0\equiv \pi m_g^2 v^2/(2g^2)$ is the kinetic energy 
associated to the initial relative velocity. 
Note that $\dot z_1 = \frac{1}{2} \dot z_{12}$ in the center of mass frame. 
This shows that the energy is exchanged between two vortices 
only through the subleading interactions in general\footnote{
The leading terms are of order $\max\{e^{-m_e a},e^{-m_g a}\}$. 
},
while the resonances dominate when $\gamma_1=\gamma_2$ and
$\gamma_1=0~(\gamma_2=0)$, 
as shown in Fig.~\ref{fig:EnergyResonance}.  

\subsection{Zero impact parameter}\label{subsec:zero}
\begin{figure}[h]
\begin{center}
\includegraphics[width=60mm]{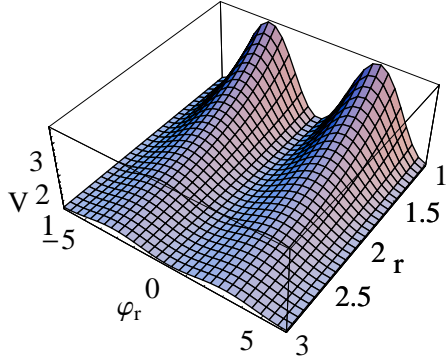}
\end{center}
\caption{The induced potential $V_Q/(g^2Q^2/8\pi)$ 
with $c_e=c_g=1.708$ and $m_e=m_g=1$.\label{fig:pot}
}
\end{figure}

In this section, we discuss the vortex scattering 
with zero impact parameter.
For simplicity, we consider the effective Lagrangian
restricted to the subspace of the moduli space given by
\beq
z_{12} = r \in \R, \hs{10} \beta_I = e^{i \varphi_I}.
\eeq
This subspace is a fixed point set of 
the following reflection symmetry on 
the moduli space
\beq
z_{12} \rightarrow \bar z_{12}, \hs{10} 
\beta_I \rightarrow \frac{1}{\bar \beta_I}.
\eeq
According to the principle of symmetric criticality, 
any solution of the equation of motion restricted on 
such an invariant submanifold is automatically 
a stationary point of the original effective action. 
On this submanifold, the Lagrangian becomes
\beq
L &=& \phantom{+} \frac{\pi}{2g^2} \left[ 1 - 2 c_e^2 K_0(m_e r) 
- 2 c_g^2 K_0(m_g r) \cos \varphi_r \right] m_g^2 \dot r^2 \notag \\
&{}& + \frac{\pi}{2g^2} \left[ 1 + c_g^2 K_0(m_g r) 
(1 + 3 \cos \varphi_r) \right] \dot \varphi_r^2 - \frac{2\pi}{g^2} 
c_g^2 K_1(m_g r) \sin \varphi_r \, m_g \dot r \dot \varphi_r \notag \\
&{}& + \frac{2\pi}{g^2} \left( 1 - 2 c_g^2 K_0(m_g r) \sin^2 
\frac{\varphi_r}{2} \right) \dot \varphi_0^2,
\eeq
where $\varphi_0 = \frac{1}{2}(\varphi_1 + \varphi_2)$ 
and $\varphi_r = \varphi_1-\varphi_2$. 
In this case, there exists only one non-vanishing conserved charge 
\beq
Q ~=~ \frac{4\pi}{g^2} \left( 1 - 2 c_g^2 K_0(m_g r) 
\sin^2 \frac{\varphi_r}{2} \right) \dot \varphi_0,
\eeq
which corresponds to the third component $(\vec Q)_3$ 
of the conserved Noether charge. 
By a Legendre transformation, 
$\dot \varphi$ can be eliminated 
from the effective Lagrangian in favor of $Q$. 
As a result, the following potential is induced 
by the Noether charge 
\beq
V_Q ~=~ \frac{g^2}{8\pi} Q^2 \left( 1 - 2 c_g^2 K_0(m_g r) 
\sin^2 \frac{\varphi_r}{2} \right)^{-1} 
~\approx~ \frac{g^2}{8\pi} Q^2 \left( 1 + 2 c_g^2 K_0(m_g r) 
\sin^2 \frac{\varphi_r}{2} \right).
\eeq
This potential is shown in Fig.~\ref{fig:pot}.
Then, the equations of motion are given by
\beq
\ddot r &=&- \left[ c_e^2 m_e K_1(m_e r) 
+ c_g^2 m_g K_1(m_g r) \cos \varphi_r \right] \dot r^2 \notag \\
&{}& - 2 c_g^2 K_0(m_g r) \sin \varphi_r \, \dot r 
\dot \varphi_r + \frac{c_g^2}{m_g} K_1(m_g r) 
\sin^2 \frac{\varphi_r}{2} \left( \frac{g^4}{4 \pi^2} Q^2 
- \dot \varphi_r^2 \right), 
\eeq
\beq
\ddot \varphi_r &=& - c_g^2 m_g^2 K_2(m_g r) \sin \varphi_r \, 
\dot r^2 + c_g^2 m_g K_1(m_g r) (1 + 3 \cos \varphi_r) \, 
\dot r \dot \varphi_r \hs{12} \notag \\
&{}& - \frac{c_g^2}{2} K_0(m_g r) \sin \varphi_r 
\left( \frac{g^4}{4\pi^2} Q^2 - 3 \dot \varphi_r^2 \right).
\eeq
If the charge $Q$ is sufficiently large 
compared to $\dot \varphi_r, \dot r$
and the relative angle $\varphi_r$ takes a generic value, 
two vortices recoil due to the potential $V_Q$, 
in which case we can continue to use the asymptotic metric. 
This phenomenon can be seen in the numerical calculation 
Fig.\,\ref{fig:test3}-(a).
On the other hand, when $Q$ is small or 
$\varphi_r \approx 0$ $(\vec n_1\cdot \vec n_2\approx 1)$, 
the vortices can closely approach along the valley of 
the potential $V_Q$, in which case we can trace the dynamics 
until the asymptotic metric becomes invalid.
Note that the cases with $Q \approx 0$ 
and $\varphi_r\approx {\rm const.}$ correspond to the 
resonance at $\gamma_1=\gamma_2$. 

\section{BPS Dyonic vortices \label{sec:4}}
\label{sc:dyonicVortex}

\subsection{Dyonic vortices from a mass deformation}

In this section we discuss the dynamics of BPS dyonic vortices 
in the $U(2)$ gauge theory. 
In order to obtain the dyonic vortices, 
we add the adjoint scalar field $\Sigma = \Sigma^0 t^a + \Sigma^a t^a$ 
and the following kinetic and mass terms 
\cite{Eto:2006pg} to the original Lagrangian Eq.\,\eqref{eq:L} 
\beq
\mathcal L_{{\rm kin}\;{\rm adj}} 
= \frac{1}{g^2}\tr 
\left[{\cal D}_\mu \Sigma \, {\cal D}^\mu \Sigma \right], 
\quad 
\mathcal L_{\rm mass} 
= \tr \left[ (\Sigma H - H M) (\Sigma H - H M)^\dagger \right],
\label{eq:mass}
\eeq
where $M$ is a $2$-by-$2$ mass matrix. 
This is the only consistent mass deformation preserving 
all supersymmetry, when embedded into the supersymmetric theory 
with eight supercharges\footnote{
Although the adjoint scalar $\Sigma$ and the mass $M$ can be triplets of 
$SU(2) \subset SO(4)_R$ in (2+1)-dimensional 
$\mathcal N = 4$ supersymmetric gauge theories, 
only one component of each triplets is relevant to the dyonic vortices. }.
The mass matrix can be written as a linear combination of 
the Pauli matrices
\beq
M = \frac{1}{2} \vec m \cdot \vec \sigma. 
\eeq 
In the following, we choose the mass to be $\vec m = (0,0,m)$. 
This mass term induces the non-vanishing VEV for the adjoint scalar
\beq
\langle \Sigma \rangle ~=~ M,
\eeq
and breaks the $SU(2)_{C+F}$ symmetry to $U(1)$ 
($U(1)_m$ from now on), for which the corresponding conserved charge 
is the component of $\vec Q$ parallel to $\vec m$. 

In this mass deformed model, 
the BPS bound for the energy density is given by
\cite{Eto:2005sw,Collie:2008za}
\beq
\mathcal E &\geq& - c \, f_{12} - i \epsilon^{ij} \p_i \tr \left[ \D_j H H^\dagger \right] + i \tr \left[ H M \D_0 H^\dagger - \D_0 H M H^\dagger \right] + \frac{2}{g^2} \p_i \tr \left[ F_{0i} \Sigma \right] .
\label{eq:dyonic_bound}
\eeq
The first two terms are the energy density of vortices 
and the third term is the conserved Noether charge for 
the unbroken $U(1)_m$ symmetry.
The forth term is the divergence of the the electric flux 
in the internal direction specified by the adjoint scalar $\Sigma$. 
We call this quantity simply ``electric charge density''.
The BPS bound is saturated if Eq.\,\eqref{eq:BPS} and 
the following equations are satisfied
\beq
&\displaystyle \D_0 H = - i(\Sigma H - H M), \hs{10} 
\D_i \Sigma = F_{0i}, \hs{10} \D_0 \Sigma = 0,
& \phantom{\Big[} \label{eq:dyonic1} \\
&\displaystyle \D_i \left[ \frac{2}{e^2} f_{0i} t^0 
+ \frac{2}{g^2} F_{0i}^a t^a \right] 
= - i ( H \D_0 H^\dagger - \D_0 H H^\dagger ).
& 
\label{eq:dyonic2}
\eeq
These BPS equations for dyonic vortices can be simplified 
by choosing the gauge in which the time component of 
the gauge field takes the form 
\beq
W_0 = M - \Sigma.
\eeq
Then, Eqs.\,\eqref{eq:dyonic1}, \eqref{eq:dyonic2} become
\beq
&\displaystyle \p_0 H = i [H, M], \hs{10} \p_0 W_i = i [W_i,M], 
\hs{10} \p_0 \Sigma = i[ \Sigma, M],& \phantom{\Big[} 
\label{eq:dyonic3} \\
&\displaystyle  \D_i \D_i \left[ \frac{2}{e^2} \Sigma^0 t^0 
+ \frac{2}{g^2} \Sigma^a t^a \right] - \{ HH^\dagger, \Sigma \} 
~=~ - 2 H M H^\dagger.& \label{eq:dyonic4}
\eeq
From these equations we find that 
the BPS dyonic vortex solutions can be obtained by solving 
Eq.\,\eqref{eq:dyonic4} with respect to $\Sigma$ 
in a static vortex background satisfying Eq.\,\eqref{eq:BPS} 
and then rotating the orientation as
\beq
H \rightarrow U^\dagger H U, \hs{10} 
W_i \rightarrow U^\dagger W_i U, \hs{10} 
\Sigma \rightarrow U^\dagger \Sigma U, \hs{10} 
U = e^{i M t} \in U(1)_m.
\eeq
Therefore, the dyonic vortex solutions are 
stable stationary configurations with rotating orientation. 
An important fact is that the no static force is exerted 
among the BPS dyonic vortices, that is, 
the electrostatic interaction Eq.\,\eqref{eq:QQ} is canceled by 
another interaction induced by the adjoint scalar $\Sigma$.

\begin{figure}[h]
\begin{center}
\begin{tabular}{ccc}
\includegraphics[width=55mm]{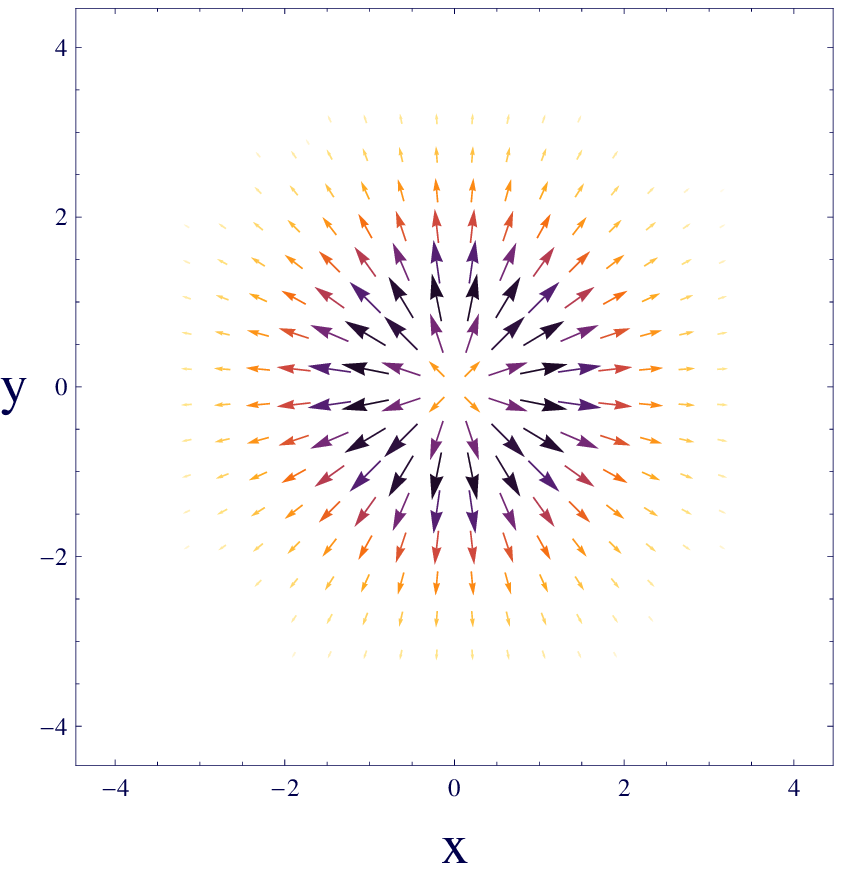} & 
\includegraphics[width=55mm]{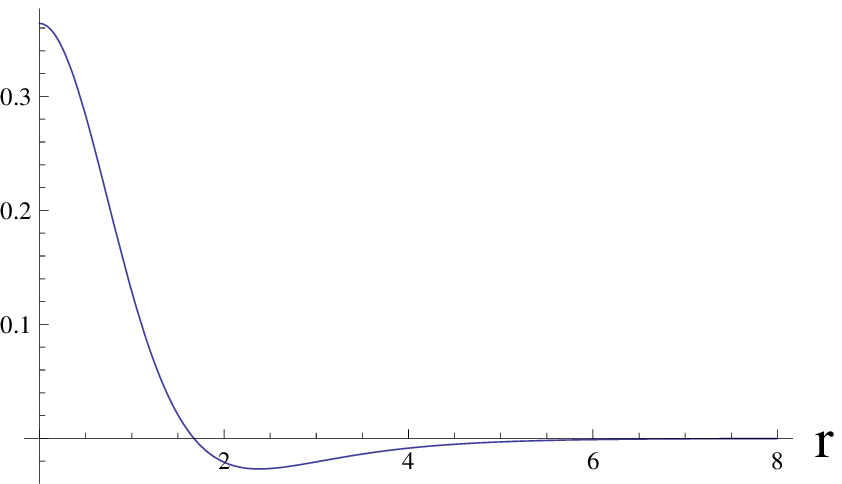} & 
\includegraphics[width=55mm]{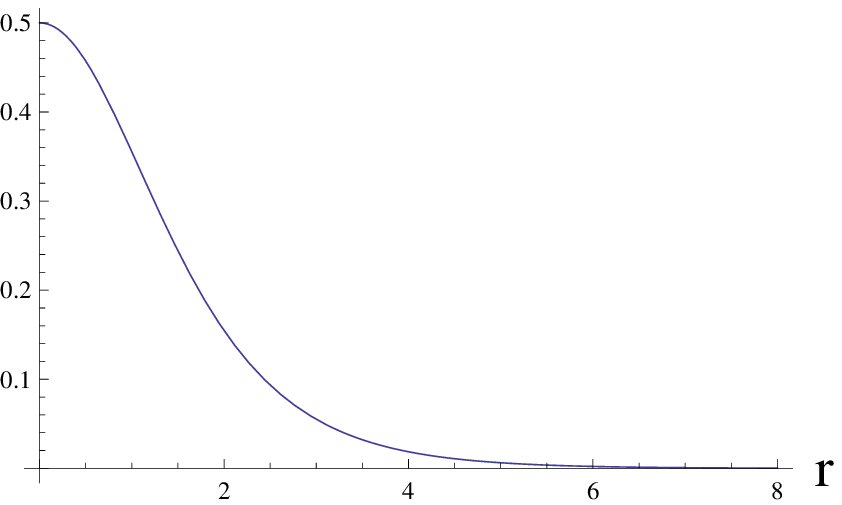} \\
(a) $\Tr [ F_{0i} \Sigma ]$ &
(b) $\p_i\Tr [ F_{0i} \Sigma ]$ & 
(c) $i \Tr[HM \D_0 H^\dagger - (c.c.)]$
\end{tabular}
\end{center}
\caption{An example of a single dyonic vortex configuration (numerical solution) for $m_g=m_e=m=1$,~$\theta_0=\frac{\pi}{2}$. (a) The electric flux radially spreads from the vortex center and decays exponentially. (b) The electric charge density is positive inside the vortex core while it is negative around the core. (c) The conserved Noether charge density is positive everywhere and the total charge is non-zero. }
\label{fig:charge}
\end{figure}

One can find the energy of the BPS saturated configurations 
by integrating the right-hand side of Eq.\,\eqref{eq:dyonic_bound} as
\beq
E = 2 \pi c k + \vec m \cdot \vec Q,
\eeq
where we have used
\beq
\vec m \cdot \vec Q = i \int d^2 x \, \Tr \left[ H M \D_0 H^\dagger - \D_0 H M H^\dagger \right].
\eeq
Note that the forth term in Eq.\,\eqref{eq:dyonic_bound}, 
as well as the second term, 
has no contribution to the total energy  
since the electric field is screened and 
decays exponentially in the Higgs phase. 
Note that the dyonic vortices have non-trivial charge distributions 
(see Fig.\,\ref{fig:charge}), although they have no total electric charge. 

We can also discuss the dyonic configurations 
in the effective theory of vortices. 
The deformation term Eq.\,\eqref{eq:mass} induces 
the following potential on the moduli space
\beq
V_m = g_{i \bar j} k^i \bar k^j, \hs{10} k^i \equiv \vec m \cdot \vec \xi^i,
\eeq
where $\vec \xi^i $ is the holomorphic Killing vector for
$SU(2)$ defined in Eq.(\ref{eq:Killing}) 
and $k^i$ is that for $U(1)_m$. 
In this deformed effective theory of vortices, 
the energy is bounded by the Noether charge $Q$ 
\beq
E ~=~ g_{i \bar j} \left( \dot \phi^i \dot{\bar \phi}^j 
+ k^i \bar k^j\right) 
~=~ g_{i \bar j} \left[ \left( \dot \phi^i \mp k^i \right)
\overline{\left( \dot \phi^j \mp k^j \right)} 
\pm \left( \dot \phi^i \bar k^j + \dot{\bar \phi}^j 
k^i \right) \right] ~\geq~\pm \vec m \cdot \vec Q.
\label{eq:dyononic_bound}
\eeq
Therefore the solutions of the dyonic vortices, 
which are trajectories on the moduli space 
with minimum energy for a given value of $\vec m \cdot \vec Q$, 
are determined from the BPS equation
\beq
\dot \phi^i = \pm  k^i.
\label{eq:dyonic}
\eeq
In the rest of this paper, we take the positive sign for 
BPS dyonic vortices. 
In the case of $\vec m = (0,0,m)$, the dyonic vortex 
solution is given by 
\beq
z_I = z_{I0}, \hs{5} \beta_I = \beta_{I0} e^{i m t}, \hs{10} 
\left( \because ~ k^i \frac{\p}{\p \phi^i} 
= i m \sum_{I=1}^k \beta_I \frac{\p}{\p \beta_I} \right),
\eeq
where $z_{I0}$ and $\beta_{I0}$ are complex constants 
which are related to the Noether charge by
\beq
\vec m \cdot \vec Q ~=~ g_{i \bar j} (\dot \phi^i \bar k^j 
+ \dot{\bar \phi}^j k^i) ~=~ 2 g_{i \bar j} k^i \bar k^j,
\eeq
where we have used Eq.\,\eqref{eq:dyonic}. 
Again, we find that the orientations of the vortices 
are rotating with the same period.

\begin{figure}[h]
\begin{center}
\includegraphics[width=40mm]{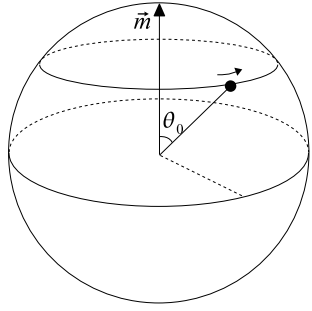}
\end{center}
\caption{The trajectory of the internal orientation
for a single dyonic vortex. 
The orientation rotates around a small circle at a fixed latitude $\theta_0$.}
\label{fig:dyonic}
\end{figure}
As an example let us consider the case of a single dyonic vortex. 
Since the dynamics of the vortex position is trivial in this case, 
we consider only the vortex orientation. 
For a single vortex, the mass deformed effective Lagrangian 
can be written in terms of the spherical coordinates 
$\beta = \tan \frac{\theta}{2} e^{i \varphi}$ as 
\beq
L = \frac{\pi}{g^2} \left[ \dot \theta^2 + \sin^2 \theta 
( \dot \varphi^2 - m^2 ) \right].
\eeq
The Noether charge $Q \equiv \vec m \cdot \vec Q/|\vec m|$ 
in this case is the conjugate momentum of $\varphi$
\beq
Q ~=~ \frac{\p L}{\p \dot \varphi} 
~=~ \frac{2\pi}{g^2} \sin^2 \theta \, \dot \varphi. 
\eeq
The dyonic 
vortex solution is given by
\beq
\theta = \theta_0, \hs{10} \varphi = m t + \varphi_0, \hs{10} 
Q =  \frac{2\pi m}{g^2} \sin^2 \theta_0.
\label{eq:dyonicVortex}
\eeq
where $\theta_0$ and $\varphi_0$ are constants. 
The trajectory corresponding to the dyonic vortex is 
a small circle at $\theta_0$, 
on which the following effective potential is minimized
\beq
V(\theta) ~=~ V_Q + V_m ~=~ \frac{g^2}{4\pi \sin^2 \theta} Q^2 
+ \frac{\pi \sin^2 \theta}{g^2} m^2 
~\geq~ \frac{2\pi m^2}{g^2} \sin^2 \theta_0,
\eeq
where $V_Q$ is the potential induced by the Noether charge $Q$. 
Let us consider a small 
fluctuation $\delta \theta$ from the trajectory 
(\ref{eq:dyonicVortex}) corresponding to the dyonic vortex: 
$\theta = \theta_0 + \delta \theta$. 
The effective potential for $\delta \theta$ is given by
\beq
V(\theta) ~\approx~ \frac{2\pi m^2}{g^2} \sin^2 \theta_0 
+ \frac{4\pi m^2}{g^2} \cos^2 \theta_0 \, \delta \theta^2.
\label{eq:massiveFluctuation}
\eeq
Therefore, if the dyonic vortex is excited, 
the angle parameter $\theta$ oscillates 
with frequency $\omega = 2 m \cos \theta_0$,
which is generically of order $m=|\vec m|$. 
Thus we find that there exist massive modes 
around the dyonic vortex configuration.

\subsection{Dynamics of dyonic vortices} 

In this section we discuss the dynamics of two dyonic vortices 
by using the equations of motion Eqs.\,\eqref{eq:eomz12} and \eqref{eq:eomq1} 
modified by the mass term. 
Since the action of the Killing vector on the moduli parameters 
is given by
\beq
(\vec m \cdot \vec \xi+\vec m\cdot \vec \xi^*) \, z_I = 0, \hs{10} 
(\vec m \cdot \vec \xi+\vec m\cdot \vec \xi^*) \, \vec n_I = \vec m\times
\vec n_I,
\eeq 
the Killing potential $V_m$ can be obtained from 
the kinetic terms of the effective Lagrangian by dropping $\dot z_I$ 
and replacing $\dot{\vec n}_I$ as
\beq
\dot{\vec n}_I \to \vec m \times \vec n_I.
\eeq
The full mass deformed Lagrangian can be found 
in Appendix \ref{eq:fullLagrangian}.  
Then the BPS equations for dyonic vortices can be rewritten as 
\begin{eqnarray}
\dot{\vec n}_I 
= \, \vec m \times \vec n_I.
\label{eq:bps_ori_m}
\end{eqnarray}
The BPS solution takes the form
\beq
\vec n_I = \frac{1}{m} \left[ \sqrt{1 - |\vec v_I|^2} \, \vec m + \cos (m t) \, m \vec v_I + \sin (mt) \, \vec m \times \vec v_I \right], 
\eeq
where $\vec v_I$ are vectors such that $\vec v_I \cdot \vec m = 0$.

Let us next see how the equations of motion for the relative position 
and the orientations are modified. 
The mass deformation changes $\vec \alpha_1 \cdot \vec \alpha_2$ 
in the last term of the equation of motion Eq.\,(\ref{eq:eomz12}) as
\begin{eqnarray}
\vec \alpha_1 \cdot \vec \alpha_2
\quad \to \quad  
\vec \alpha_1 \cdot \vec \alpha_2
-\vec \alpha'_1 \cdot \vec \alpha'_2,
\label{eq:spatialEOMchange}
\end{eqnarray}
where $\vec \alpha'_I$ is the vectors obtained from 
$\vec \alpha_I = \dot{\vec n}_I - i \vec n_I \times \dot{\vec n}_I$ 
by replacing $\dot {\vec n}_I$ with $\vec m \times \vec n_I$ 
\begin{eqnarray}
\vec \alpha'_I ~\equiv~ 
\vec m \times \vec n_I - i \vec n_I \times (\vec m \times \vec n_I).
\end{eqnarray}
Therefore, if the pair of the vortices are near BPS 
$\dot{\vec n}_I \approx \vec m \times \vec n_I$, 
the contributions from the last term of Eq.\,(\ref{eq:eomz12}) and
the mass deformation are small 
\beq
\vec \alpha_1 \cdot \vec \alpha_2
-\vec \alpha'_1 \cdot \vec \alpha'_2 ~\approx~ 0.
\eeq 
Similarly, we can also show that there exists a similar cancellation 
in the equation of motion for the orientation Eq.\,\eqref{eq:eomq1}
\beq
\vec \alpha_1 (\vec n_2 \cdot \vec \alpha_1 + 2 \vec n_1 \cdot \vec \alpha_2 ) 
-
\vec \alpha_1' (\vec n_2 \cdot \vec \alpha_1' + 2 \vec n_1 \cdot \vec \alpha_2' ) ~\approx~ 0.
\eeq 
In the following, we will see that due to these cancellation, 
the behavior of the dyonic vortices are quite different 
from that of vortices before the mass deformation. 

\subsubsection{Angular momentum, Lorentz Force and a Bound State}
Let us first discuss the case in which 
the relative velocity of the dyonic vortices are small. 
By using the effective Lagrangian Eq.\,(\ref{eq:eff-Lag}), 
we can write down the conserved angular momentum 
in terms of the radial coordinates $z_{12}=r e^{i\chi}$ as
\begin{eqnarray}
{\bf L}_\chi ~=~ \frac{\partial L}{\partial \dot \chi}
~\approx~ \frac{\pi m_g^2}{g^2} r^2 \, \dot \chi + A_\chi(r), 
\label{eq:angular}
\end{eqnarray}
where we have neglected a small term proportional to $\dot \chi K_0$. 
The function $A_{\chi}$ is given by
\beq
A_\chi(r) = - \left( \vec n_1 \cdot \vec Q_2 + \vec n_2 \cdot \vec Q_1 \right)
c_g^2 K_1(m_g r) m_g r.
\eeq
The first term in Eq.\,\eqref{eq:angular} is the ordinary angular momentum of 
a free particle and the additional term $A_\chi(r)$ can be interpreted 
as a ``gauge potential'' generated 
by the motion of the orientational moduli. 

\begin{figure}[h]
\begin{center}
\begin{tabular}{ccc}
\includegraphics[width=70mm]{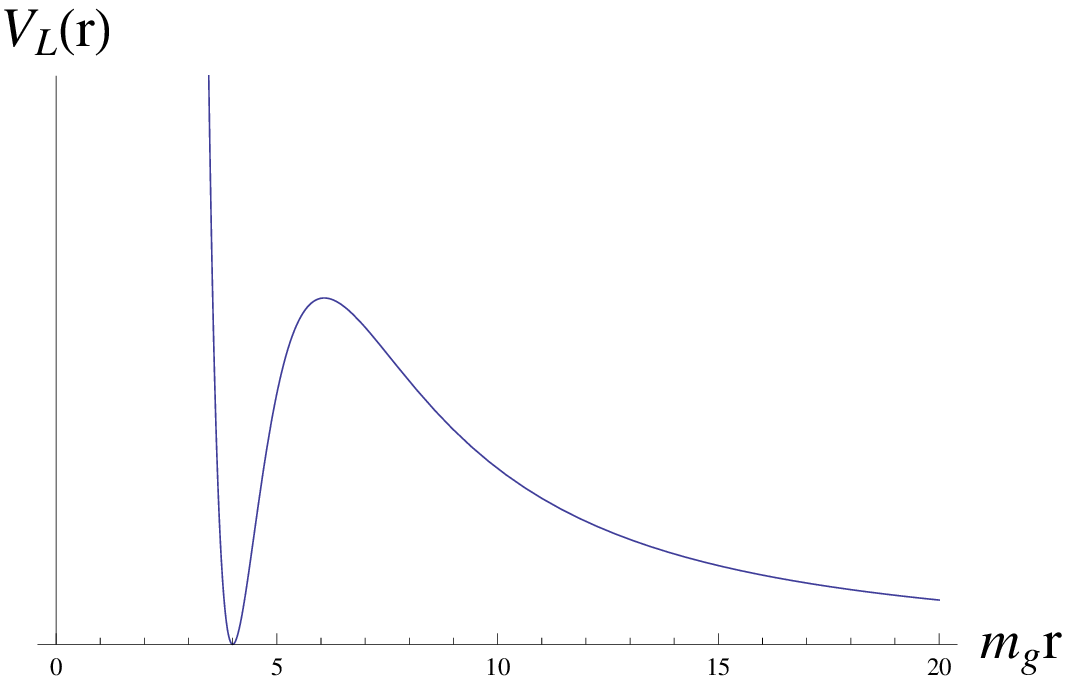} &\qquad & 
\includegraphics[width=70mm]{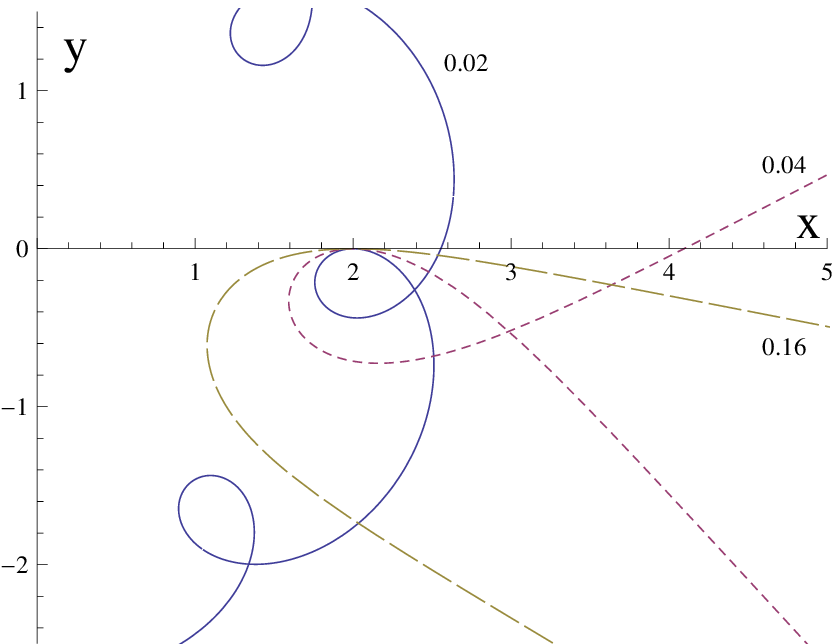}\\
(a)&&(b)
\end{tabular}
\caption{
(a) The potential $V_L(r)$ representing the Lorentz force for $m_g r_0=4$.  
(b) Examples of orbits (numerical solutions) for 
$m_g v / |\vec m | = 0.02$ (solid line), $0.04$ (dotted line), 
$0.16$ (dashed line). 
Solid line shows an extended coil winding circularly around the origin. }
\label{fig:VL}
\end{center}
\end{figure}

\begin{figure}[h]
\begin{center}
\begin{tabular}{ccc}
\includegraphics[width=50mm]{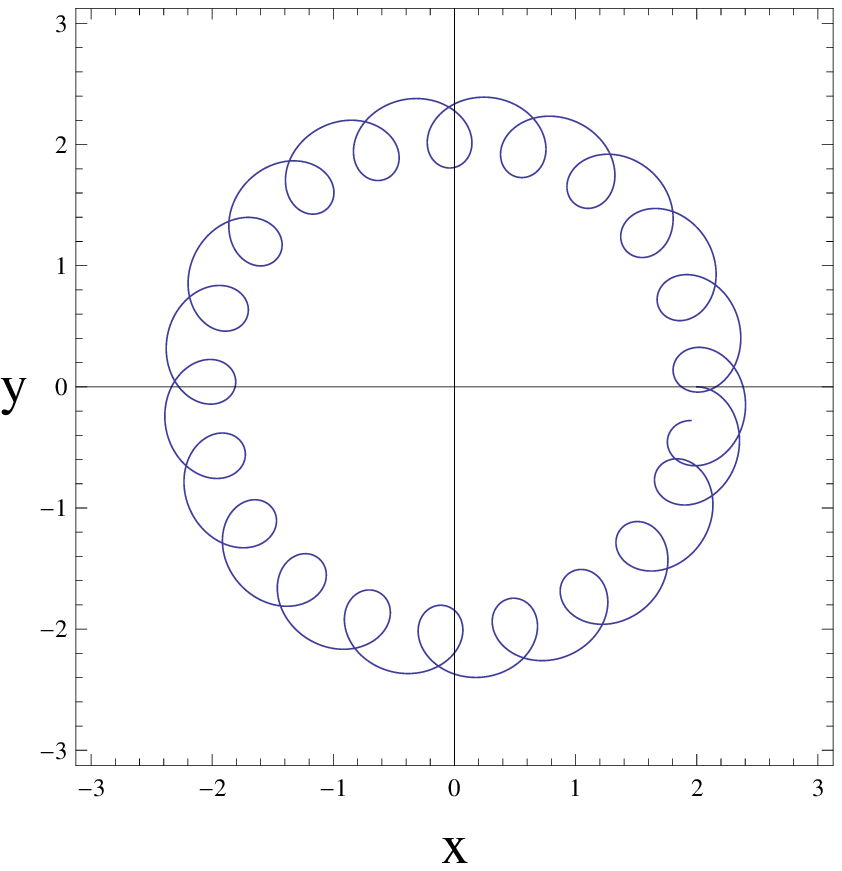} &\qquad &
\includegraphics[width=70mm]{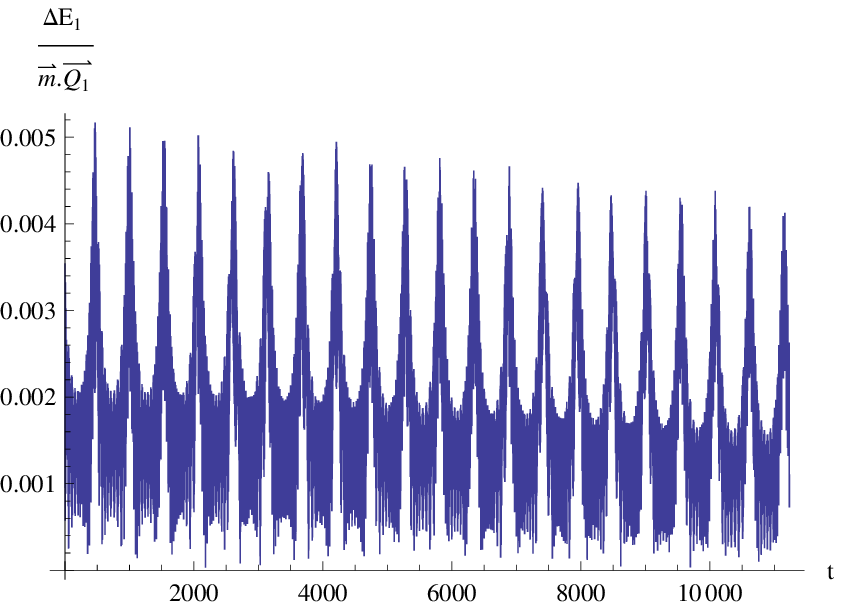} \\
(a) Orbit of an extended coil around a circle
&& 
(b) Deviation of energy from the BPS mass 
\end{tabular}
\caption{\small 
An example of the bound state (numerical solution) 
for $m_g v/|\vec m|=0.016$. 
(a) The orbit in $z$-plane.  
(b) The deviation of energy $\Delta E_1$ 
(divided by the BPS mass) as a function of time. 
}
\label{fig:boundstate}
\end{center}
\end{figure}

In the massless case discussed in section \ref{sec:3}, 
the contribution of $A_\chi(r)$ is small compared with the leading terms.
This can be seen by averaging over the rapid motions of the orientations. 
Since they rotate around great circles of $\C P^1$, it follows that
\beq
\left< \vec n_1 \cdot \vec Q_2 + \vec n_2 \cdot \vec Q_1 \right> = 0.
\eeq
The mass deformation drastically changes this situation. 
Since the orientations rotate around small circles 
in the case of the dyonic vortices, 
the average of the gauge potential is non-zero
\begin{eqnarray}
\left< \vec n_1 \cdot \vec Q_2 + \vec n_2 \cdot \vec Q_1 \right>
= \frac{2\pi}{g^2}\,
\vec m \cdot ( \vec n_1 + \vec n_2 )( 1 - \vec n_1 \cdot \vec n_2).
\end{eqnarray}
Note that $\vec m \cdot {\vec n}_I$ and $\vec n_1 \cdot \vec n_2$ 
are independent of time for BPS dyonic vortices and 
slowly vary for an interacting pair of vortices. 
Due to the contribution from the gauge potential $A_\chi$, 
the motions of the dyonic vortices 
with small relative velocity $v = |\dot z_{12}|$
becomes quite different from that without the mass deformation.
For a pair of near BPS dyonic vortices 
$\dot{\vec n}_I \approx \vec m \times \vec n_I$, 
the non-trivial gauge potential $A_\chi(r)$ 
gives the dominant contribution via the Lorentz force 
\begin{eqnarray}
\frac{\pi m_g^2}{g^2} \ddot z_{12} \ \approx \ -i B \, \dot z_{12}, \qquad  
B(r)\equiv \frac1r \frac{d A_\chi(r)}{dr}.
\label{eq:Lorentz}
\end{eqnarray}
Assuming that the relative velocity $v = |\dot z_{12}|$ and 
the massive oscillations 
[$\delta \theta$ in Eq.\,(\ref{eq:massiveFluctuation})] are small, 
we can write down the energy of this system as
\begin{eqnarray}
E ~\approx~ \frac{\pi m_g^2}{2g^2} \dot r^2 + 
V_{\rm L}(r), \qquad 
V_{\rm L}(r) = \frac{g^2}{\pi} 
\left( \frac{ \mathbf L_{\chi} - A_\chi(r) }{m_gr}\right)^2,
\end{eqnarray}
where we have neglected irrelevant terms. 
For an arbitrary relative distance $r=r_0$, 
we can always adjust $\mathbf L_{\chi}$ so that $V_{\rm L} = 0$, 
namely
\beq
\mathbf L_{\chi} = A_{\chi}(r_0).
\eeq 
This minimum energy configuration corresponds to the BPS solution, 
for which the velocity of the relative position is zero 
$\dot r = \dot \chi = 0$.
As shown in Fig.\,\ref{fig:VL}-(a), 
this is a stable point of the positive semi-definite potential $V_{\rm L}$, 
and thus the relative distance oscillates around the minimum 
for a sufficiently small excitation energy.
The time dependence of the relative angle $\chi$ can be determined from 
Eq.\,\eqref{eq:angular}.
As a result, we can find that the dyonic vortices drift 
along a contour line of the magnetic field
similarly to a charged particle in a background magnetic field. 
The orbit of each dyonic vortex in the $z$-plane 
takes the form of an extended coil winding around a contour line 
(see Fig.\,\ref{fig:boundstate}-(a)). 
In this sense, a pair of dyonic vortices can form 
a ``bound state" with a large relative distance $r_0 \gg m_g^{-1}$. 
The frequency $\omega_{\rm L}$ of the oscillation around 
the stable point $r=r_0$ is given by
\beq
\omega_{\rm L} ~\approx~ \frac{g^2}{\pi m_g^2} B(r_0) ~\propto~ K_0(m_g r_0),
\eeq
and is generically quite small $|\omega_{\rm L}|\ll |\vec m|$ 
for well-separated vortices with $m_g r_0 \gg 1$. 
The radius of the coil $r_{\rm coil}$, 
the velocity of the drift $v_{\rm D}$ and 
the period of the large circular motion $T_B$ are respectively estimated as
\beq
r_{\rm coil} ~\approx~ \frac{v}{|\omega_{\rm L}|}, \hs{10} 
v_{\rm D} ~\approx~ \frac{m_g v^2}{2|\omega_{\rm L}|}, \hs{10}
T_B ~\approx~ \frac{4 \pi |\omega_{\rm L}| r_0}{m_g v^2}, 
\eeq
where we have assumed that the relative velocity 
$v = |\dot z_{12}|$ is sufficiently small.
Note that the dyonic vortices with $v$ larger than 
a certain critical value run away to infinity 
as illustrated in Fig.\,\ref{fig:VL}-(b). 

In this analysis, we have neglected subleading terms 
in the equation of motion. 
This is valid if the excitation energy is sufficiently small 
since the subleading terms just change the potential $V_{\rm L}(r)$ slightly 
for a small relative velocity $v$. 
We have also assumed that the motions of the orientations are kept near BPS 
$\dot n_I \approx \vec m \times \vec n_I$ 
and fluctuations of the massive modes $\delta \theta_I$ are small. 
These assumptions are justified if we take 
the near BPS initial condition with sufficiently small $v$, 
since the geodesic force which is independent of $\dot z_{12}$ 
is almost cancelled by the potential term induced by the mass deformation.
Therefore, we expect that the bound state is stable 
as long as the deviation of the energy of each vortex is 
sufficiently small. 
Fig.\,\ref{fig:boundstate}-(b) shows an example of 
the deviation of the energy from the BPS mass 
during the period $T_{\rm B}$.  

Now let us discuss the origin of the Lorentz force Eq.\,\eqref{eq:Lorentz}. 
Since dyonic vortices have both magnetic and electric fluxes, 
their asymptotic interaction can be well described by regarding them 
as point-like sources with both magnetic and electric moments. 
The static electric-electric and magnetic-magnetic interactions 
are canceled by the other scalar fields, 
so that the dominant interaction for a small relative velocity is 
the electric-magnetic interaction which is proportional to $\dot z_{12}$. 
Therefore, each slowly moving dyonic vortex feels 
the Lorentz force of the form Eq.\,\eqref{eq:Lorentz} 
from the fluxes of the other dyonic vortex.

\subsubsection{Scattering of dyonic vortices}

In this section, we discuss the scattering of dyonic vortices 
by using the free motion approximation as in section \ref{sec:3}. 
For simplicity, we restrict the initial conditions for the vortices 
to the BPS states, that is
\beq
z_{12}= vt + i a, \hs{10} 
\vec n_I = \frac{1}{m} \left[ \sqrt{1 - |\vec v_I|^2 } \vec m + \cos (m t) \, \vec v_I + \sin ( m t ) \, \vec m \times \vec v_I \right].
\eeq
Substituting this into the equation of motion 
Eq.\,\eqref{eq:eomz12} modified by the mass deformation
Eq.\,(\ref{eq:spatialEOMchange}), 
we obtain the following equation for the relative position $z_{12}$
\beq
\ddot z_{12} &=& - m_e v^2 c_e^2 
\frac{v t - i a}{\sqrt{(v t)^2 + a^2}} K_1 
\big( m_e \sqrt{(v t)^2 + a^2} \big) \notag \\
&{}& - m_g v^2 c_g^2 \frac{v t - i a}{\sqrt{(v t)^2 + a^2}} 
K_1 \big(m_g \sqrt{(v t)^2 + a^2} \big) \, 
\vec n_{1} \cdot \vec n_{2} \notag \\
&{}& - 2 v i c_g^2 K_0 \big( m_g \sqrt{(vt)^2 + a^2} \big) \, 
\vec m \cdot ( \vec n_{1} + \vec n_{2}) 
(1 - \vec n_{1} \cdot \vec n_{2} ). \phantom{\Big[}
\label{eq:dyonic_z12}
\eeq
Here $\vec n_1 \cdot \vec  n_2$ and 
$\vec m\cdot \vec n_I$ are constant 
since we have assumed that the orientation modes satisfy the BPS equations 
$\dot{\vec n}_I = \vec m \times \vec n_I$.
Note that the forces vanish in the limit $v \rightarrow 0$ 
due to the BPS properties of the dyonic vortices. 
The total change in the relative velocity $\Delta \dot z_{12}$ is
\beq
\frac{\Delta \dot z_{12}}{2 \pi i v} = \frac{1}{2} c_e^2 e^{-m_e a} +  
\frac{1}{2} c_g^2 e^{-m_g a}  \vec n_{1} \cdot \vec n_{2} - 
c_g^2 e^{-m_g a}\frac{\vec m \cdot( \vec n_{1} + \vec n_{2} )}{m_g v} (1 - \vec n_{1} \cdot \vec n_{2}).
\label{eq:dyononic_z12}
\eeq
Similarly, we can calculate the total change in the $Q$-charge 
\beq
\Delta (\vec m \cdot \vec Q_1) \ = \ - \Delta (\vec m \cdot \vec Q_2) \ = \ - \frac{\pi c_g^2 m_g v}{2} \, (\vec n_{1} \times \vec n_{2}) \cdot \vec m \left[ 1 + \frac{2\vec m \cdot (\vec n_{1} + \vec n_{2})}{m_g v} \right] e^{-m_g a}.
\label{eq:dyononic_Q}
\eeq
From these results, we find that 
$\Delta \dot z_{12}$ and $\Delta (\vec m \cdot \vec Q_I)$ 
do not vanish even for small relative velocity\footnote{
Although the free motion approximation 
would not be a good approximation for $v \approx 0$, 
the results Eqs.\,\eqref{eq:dyononic_z12} and \eqref{eq:dyononic_Q} 
are qualitatively correct if the impact parameter $a$ is sufficiently large.}
$v \approx 0$. 
This is because the Lorentz force 
(the last term in Eq.\,\eqref{eq:dyonic_z12})
is proportional to $v$ and 
its integrated effect is independent of $v$.
Before the mass deformation, 
the differences $\Delta \dot z_{12}$ and $\Delta \vec Q_I$ 
vanish for a pair of slowly moving 
BPS vortices ($v \rightarrow 0$, $\gamma_I =0$) 
as we can see from Eqs.\,\eqref{eq:z12} and \eqref{eq:Q1}.
Note that in both cases the interaction vanish 
in the limit $v \rightarrow 0$ because of the BPS properties. 
This is one of the typical differences between BPS vortices 
before and after the mass deformation.

Since the first two terms in Eq.\,(\ref{eq:dyononic_z12})
can be obtained by taking the limit 
$\gamma_{1,2} \to 0$ in Eq.\,(\ref{eq:dz12}), 
they consistently reduce to Eq.\,(\ref{eq:dz12}) 
in the limit $|\vec m| \to 0$. 
Note that $\gamma_I \propto |\vec Q_I|$
are of order $|\vec m|$ for the BPS dyonic vortices.
The effect of the mass 
deformation lies in the last term, which 
becomes dominant if $m_g v \ll |\vec m| $.  
On the other hand, the free-motion approximation 
gives quantitatively precise results when the relative velocity 
$v$ is sufficiently large : $m_g v \gg m e^{-m_g a}$. 
In such situations\footnote{
In the opposite case $m_g v \gg |\vec m|$, 
the approximation is always valid, 
but the effects of the mass deformation become subleading. 
For instance, there is no minimum in the potential 
$V_{\rm L}(r)$ in such a case.}, 
we can show that the impact parameter 
$a$ should be sufficiently larger than 
the minimum of the potential $r_0$ which is given by
\begin{eqnarray}
|A_\chi (r_0)| ~=~ |{\bf L}_\chi| ~=~ \frac{\pi}{g^2}(m_g a) (m_g v) . 
\end{eqnarray}
Therefore Eqs.~(\ref{eq:dyononic_z12}) and 
(\ref{eq:dyononic_Q}) are valid 
only for the scattering of dyonic vortices 
repelled by the outer barrier of the potential $V_{\rm L}(r)$. 

As we can see from Eqs.\,\eqref{eq:dyononic_z12} and \eqref{eq:dyononic_Q}, 
there is no contribution of order $m^2 \propto (\gamma_I)^2$. 
This is one of the most striking features of the dyonic vortices, 
which makes the Lorentz force dominant and 
ensures the stability of the bound states. 
Strictly speaking, we have to consider contributions 
from the massive oscillations which we have ignored for simplicity. 
We expect that massive modes are irrelevant for the near BPS configurations, 
while they would cause instabilities of highly excited bound states\footnote{ 
In the latter case, there would be corrections with resonant behaviors 
in the scattering angle $\Delta \dot z_{12}$, etc. }.

\section{Conclusion and Discussion \label{sec:5}}
We have studied the dynamics 
of non-Abelian vortices by the moduli space approximation
using the asymptotic metric of well-separated non-Abelian vortices 
in the $U(N)$ gauge theory with $N$ Higgs scalar fields 
in the fundamental representation. 
Since non-Abelian vortices  
carry the orientational moduli $\mathbb CP^{N-1}$ individually, 
they can have Noether charges as internal momentum of them, 
whereas ANO vortices in the Abelian-Higgs model have no such charges.

We have found that the vortices with the same charges repel 
while those with the opposite charges attract. 
We have shown that the charges of vortices 
can change during the scattering process 
with the total energy conserved; 
the kinetic energy of orientational moduli can be 
transfered to the kinetic energy in real space and vice versa. 
As results, we have found that in scattering of two 
non-Abelian vortices, 
i) the scattering angle depends on the internal orientation, 
especially parallel orientations give repulsion 
while anti-parallel orientations give attraction, 
ii) the energy of real and internal spaces can be transfered,
iii) the energy and charge transfer between two vortices occur, 
and 
iv) some resonances appears 
due to synchronization of the orientations.

By introducing the mass deformation into the original theory, 
the color-flavor symmetry $SU(N)$ is explicitly broken to $U(1)^{N-1}$.
Noether charges of these $U(1)^{N-1}$ can be actually regard as 
$U(1)^{N-1}$ gauge charges where 
the Cartan subgroup of $U(N)$ survives in the low energy 
theory in the large mass limit. 
We have shown that the 
dominance of the Lorentz force between vortices 
gives a bound state with a coiling orbit.

\bigskip
Here we address several discussions. \\
One may question what are the conditions for 
the existence of bound states in more general cases.  
To form a bound state, therefore, vortices are needed only to have
\begin{enumerate}
\item a magnetic flux (a topological charge) as the definition,
\item a sufficiently large conserved charge,
\item a sufficiently small static force.
\end{enumerate}
Since large conserved charges generically 
induce large interactions between vortices, 
we need to prepare a special 
system for cancellation of static forces
like a BPS state.
Even in the Abelian case, vortices can have conserved charges.
For instance, Abelian semi-local vortices 
\cite{Vachaspati:1991dz,Leese:1992fn} 
can acquire $Q$-charges  
after the mass deformation. 
In the strong coupling limit $e \to \infty$, 
they reduce to $Q$-lumps 
and their interactions 
have been studied \cite{Ward:1985ij,Leese:1991hr}.

As future problems, the followings might be interesting.

Our study in this paper has been based on the moduli space metric for 
well-separated vortices and therefore we cannot discuss vortices 
with a small separation. 
One question is 
what the fate of a pair of vortices with anti-parallel charges is. 
Will they form a bound state with a very small separation? 
Or will they reduce to a composite state of 
coincident vortices (singlet or triplet) \cite{Eto:2006cx,Eto:2010aj} 
after radiations of massive particles? 
The head-on collision of two vortices was studied previously 
based on the metric on the moduli subspace of
two coincident vortices (at zero distance) \cite{Eto:2006db}. 
However the moduli space metric for two vortices at arbitrary distance 
is not known, which is needed to answer the above questions.
Also, beyond the moduli space approximation, one has to study 
dynamics by numerically solving the original equations of motion.

We have found that energy transfer and Noether charge transfer 
between two scattering vortices. 
This property may support the Boltzman's principle of equality 
when one considers statistical mechanics of many vortices. 
Partition function of a gas of non-Abelian vortices 
was calculated in \cite{Eto:2007aw} 
with assuming the principle of equality 
so that the calculation is reduced 
to the integration over the moduli space of vortices.
Therefore, one important question 
is if a large number of scattering of vortices 
leads to the ergodic theorem 
so that one can assume the Boltzman's principle of equality.

Non-Abelian vortices are called 
semi-local if there are more flavors than the number of color. 
Non-Abelian semi-local vortices have size moduli 
and they reduces to local vortices
when the size modulus is sent to zero. 
One of characteristic properties of semi-local vortices is 
that their profile functions decay polynomially 
but not exponentially as local vortices. 
Consequently the size modulus is non-normalizable. 
Moreover the ${\mathbb C}P^{N-1}$ orientational moduli 
of a single non-Abelian semi-local vortex 
are also non-normalizable \cite{Shifman:2006kd} 
when the size modulus is non-zero, 
but they are normalizable only when the size modulus 
is zero \cite{Eto:2007yv}.
No metric can be defined for non-normalizable moduli, 
and dynamics cannot be discussed for those.
However we can discuss the dynamics of two semi-local vortices 
because relative orientations and relative size 
between them are normalizable.
In fact dynamics such as head-on collision was studied for
Abelian \cite{Leese:1992fn} and 
non-Abelian \cite{Eto:2006db} semi-local vortices.
Since semi-local vortices can be approximated by lump solutions 
at large distance compared with Compton wave length of massive particles, 
the dynamics of semi-local vortices at large distance 
can be approximated by that of lumps \cite{Ward:1985ij}. 

One interesting generalization is changing geometry from 
a flat space to geometry with non-trivial cycles such as 
a cylinder \cite{Eto:2006mz}, a torus \cite{Eto:2007aw}, 
and Riemann surfaces with higher genus \cite{Baptista:2008ex}.
For semi-local vortices in compact Riemann surfaces, 
there is no problem of the non-normalizability 
and we can discuss their dynamics 
without introducing any cut-off scale by hand.
In fact the moduli space metric was found 
for well-separated non-Abelian vortices on Riemann surfaces 
\cite{Baptista:2010rv}. 
However dynamics during a long time 
will be difficult using the asymptotic metric 
because distances between vortices cannot be kept large for compact spaces. 

Another interesting extension is changing gauge groups.
Non-Abelian vortices were extended to arbitrary gauge groups $G$  
in the form of ${(U(1) \times G)/ C(G)}$ with the center $C(G)$ of $G$
\cite{Eto:2008yi}. 
Especially the cases of $G=SO(N), USp(2N)$ have been studied in detail
\cite{Eto:2008qw,Ferretti:2007rp}. 
However they are semi-local vortices in general \cite{Eto:2008qw} 
so that dynamics at large separation is the one of lumps \cite{Ward:1985ij}.

We have studied dynamics of BPS vortices. 
There exist static force between 
non-BPS vortices \cite{Speight:1996px}.
Superconductors are classified into 
two types, type I and type II, depending on 
whether the static force between 
two vortices is attractive(type I) or repulsive (type II).
A classification of non-Abelian superconductors 
will be more complicated due to the existence of charges 
of non-BPS non-Abelian vortices \cite{Auzzi:2007wj}. 
When the couplings are close to the critical coupling 
(near BPS), dynamics of non-BPS vortices 
can be studied by the moduli space approximation 
plus a potential for the static forces, 
which remains as an important future problem.
Inclusion of Chern-Simons terms \cite{Collie:2008mx} is also 
possible extension.

Our studies have been restricted to particle dynamics, 
namely vortices in 2+1 dimensions. 
In 3+1 dimensions, vortices are (cosmic) strings 
which have one spatial dimension in their world-sheet.
The moduli space approximation can be applied 
when the angle between two cosmic strings is small.
Collision of two non-Abelian cosmic strings was studied 
based on the metric on the moduli subspace 
of two coincident vortices \cite{Hashimoto:2005hi,Eto:2006db}. 
Especially it was found in \cite{Eto:2006db}
that orientational moduli of 
two non-Abelian vortices  
must be aligned and scatter with 90 degree angle, 
when they collide in head on, 
except for a fine-tuned collision. 
This implies that two non-Abelian cosmic strings reconnect 
each other. 
This result was obtained just before and after the collision 
moment in the linear order in time. 
On the other hand, Eq.\,(\ref{eq:collision}) shows that orientational moduli of 
two vortices in head-on collision tend to be aligned. 
It suggests that two non-Abelian cosmic strings consistently reconnect each other as a long time behavior by 
feeling attraction between two orientational moduli like a ferromagnet.
However we also have found in this paper that two vortices repel when 
charges induced by the motion of internal orientational moduli of 
two vortices are opposite. 
In this case, we expect that two cosmic strings merely 
reconnect each other. 

We have found many new feature of 
dynamics or scattering of non-Abelian vortices, 
which are based on the existence of non-Abelian internal moduli 
of individual solitons. 
Therefore the similar properties should hold 
for scattering of other kinds of non-Abelian solitons 
such as Yang-Mills instantons, 
non-Abelian monopoles \cite{Goddard:1976qe}, and 
non-Abelian kinks \cite{Shifman:2003uh}.  

\section*{Acknowledgments}

The work of M.N.~and of N.S~are supported in part by 
Grant-in-Aid for Scientific 
Research No.~20740141 (M.N.), No.~21540279 (N.S.) and 
No.~21244036 (N.S.) from the Ministry 
of Education, Culture, Sports, Science and Technology-Japan. 
One of the authors (N.S.) would like to thank Nick Manton 
and David Tong for a useful discussion. 

\appendix

\section{The action in terms of the orientation vectors $\vec n_I$ 
\label{eq:fullLagrangian}}
In the case of $U(2)$ gauge theory,  
it is convenient to describe orientations of vortices 
in terms of three-component unit vectors $\vec n_I$.
The K\"ahler potential in Eqs.~(\ref{eq:freeKahlerpotential}) 
and (\ref{eq:Kahlerpotential}) 
give the effective Lagrangian 
\begin{eqnarray}
L = L_{\rm kin} + L_{\rm int} - V_{m}. 
\label{eq:eff-Lag}
\end{eqnarray}
Introducing Lagrange multipliers  $\lambda_I\in \mathbb R$, 
the effective Lagrangian ${\cal L}_{\rm eff}$ can be rewritten 
in terms of $\vec n_I$.
The kinetic terms giving free motions are obtained as
\begin{eqnarray}
L_{\rm kin} = \sum_{I} \frac{2\pi}{g^2} \left(\frac{1}{2} 
m_g^2 |\dot z_I|^2 + \frac{1}{2} |\dot {\vec n}_I|^2 
+\lambda_I(|\vec n_I|^2-1)\right).
\end{eqnarray}
 Note that the quantity  $\Theta_{12}$ in the K\"ahler
 potential is $\Theta_{12}=\vec
 n_1\cdot \vec n_2$ and 
\begin{eqnarray}
 d\beta\frac{\partial}{\partial \beta}=
\frac12\left(d\vec n-\frac{\vec n}{|\vec n|^2}(\vec n\cdot d\vec n)
-i\vec n\times d\vec n\right)\cdot \frac{\partial}{\partial \vec n}.
\end{eqnarray}
Therefore 
kinetic terms $L_{\rm int}$ describing 
interactions between vortices can be 
directly calculated from the K\"ahler potential as 
\begin{eqnarray}
\frac{g^2}{2\pi} L_{\rm int}
&=&-\frac12m_g^2\left(\dot r^2+r^2 \dot \chi^2\right)
\left(c_e^2 K_0(m_e r)
+ c_g^2(\vec n_1\cdot {\vec n}_2) K_0(m_g r)\right) 
\nonumber \\
&&+c_g^2\left\{\left((\vec n_1\cdot \dot {\vec n}_2)+
(\vec n_2\cdot \dot {\vec n}_1)\right)m_g\dot r 
+(\vec n_1\times \vec n_2)\cdot(\dot{\vec n}_1-\dot{\vec
       n}_2)m_gr \dot \chi \right\} K_1(m_gr)\nonumber\\
&&-c_g^2 \left\{-\vec n_1\cdot \vec n_2
(|\dot {\vec n}_1|^2+|\dot {\vec n}_2|^2)+\dot{\vec n}_1 
\cdot \dot{\vec n}_2
+(\vec n_1 \times \dot{\vec n}_1)\cdot 
(\vec n_2 \times \dot{\vec n}_2)\right\}
K_0(m_g r),\qquad
\end{eqnarray}
where $z_1-z_2=z_{12}=r e^{i\chi}$. 
 The potential $V_{m}$ induced by the mass deformation 
can be obtained by just replacing $\dot z_I\to 0$ and 
$\dot n_I\to \vec m\times \vec n_I$ as
\begin{eqnarray}
\frac{g^2}{2\pi} V_{m}&=&\sum_{I}\frac12|\vec m\times \vec n_I|^2 
-c_g^2 \Big\{-\vec n_1\cdot \vec n_2
(|\vec m\times {\vec n}_1|^2+|\vec m\times {\vec n}_2|^2)
\nonumber\\
&& {}+(\vec m\times {\vec n}_1) \cdot (\vec m \times {\vec n}_2)
+(\vec n_1 \times (\vec m\times {\vec n}_1))\cdot 
(\vec n_2 \times (\vec m \times{\vec n}_2))\Big\}K_0(m_gr).
\end{eqnarray}
Due to the Lagrange multipliers, 
the equations of motion for the orientations become
\begin{eqnarray}
({\bf 1}_3-\vec n_I \vec n_I^{\rm T})
\left(\frac{\partial {\cal L}}{\partial \vec n_I}-
\frac{d}{dt}\frac{\partial {\cal L}}{\partial \dot{\vec n}_I}
\right) =0, \qquad 
\Leftrightarrow \qquad 
\vec n_I\times \left(\frac{\partial {\cal L}}{\partial \vec n_I}-
\frac{d}{dt}\frac{\partial {\cal L}}{\partial \dot{\vec n}_I}
\right) =0.
\end{eqnarray}
 For instance, the equation of motion of the 
orientation $\vec n$ for a single vortex is given by 
\begin{eqnarray}
\vec n \times \ddot{\vec n} + (\vec m\cdot \vec n) \, \vec m \times \vec n = 0,
\end{eqnarray}
with the conservation conditions $\vec n \cdot \dot {\vec n}=0$, 
$|\vec n|^2=1$.



\begin{thebibliography}{99}

\bibitem{Manton:2004tk}
  N.~S.~Manton and P.~Sutcliffe,
  ``Topological solitons,''
{\it  Cambridge, UK: Univ. Pr. (2004) 493 p}

\bibitem{Belavin:1975fg}
  A.~A.~Belavin, A.~M.~Polyakov, A.~S.~Shvarts and Yu.~S.~Tyupkin,
  ``Pseudoparticle solutions of the Yang-Mills equations,''
  Phys.\ Lett.\  B {\bf 59}, 85 (1975).

\bibitem{'tHooft:1974qc}
  G.~'t Hooft,
  ``Magnetic Monopoles In Unified Gauge Theories,''
  Nucl.\ Phys.\  B {\bf 79}, 276 (1974); 
  A.~M.~Polyakov,
  ``Particle spectrum in quantum field theory,''
  JETP Lett.\  {\bf 20}, 194 (1974)
  [Pisma Zh.\ Eksp.\ Teor.\ Fiz.\  {\bf 20}, 430 (1974)].

\bibitem{Abrikosov:1956sx}
  A.~A.~Abrikosov,
  ``On the Magnetic properties of superconductors of the second group,''
  Sov.\ Phys.\ JETP {\bf 5}, 1174 (1957)
  [Zh.\ Eksp.\ Teor.\ Fiz.\  {\bf 32}, 1442 (1957)];
  H.~B.~Nielsen and P.~Olesen,
  ``Vortex-line models for dual strings,''
  Nucl.\ Phys.\  B {\bf 61}, 45 (1973).

\bibitem{Abraham:1992vb}
  E.~R.~C.~Abraham, P.~K.~Townsend,
  ``Q kinks,''
  Phys.\ Lett.\  {\bf B291}, 85-88 (1992); 
  ``More on Q kinks: A (1+1)-dimensional analog of dyons,''
  Phys.\ Lett.\  {\bf B295}, 225-232 (1992);
  M.~Arai, M.~Naganuma, M.~Nitta, N.~Sakai,
  ``Manifest supersymmetry for BPS walls in N=2 nonlinear sigma models,''
  Nucl.\ Phys.\  {\bf B652}, 35-71 (2003).
  [hep-th/0211103]; 
  ``BPS wall in N=2 SUSY nonlinear sigma model with Eguchi-Hanson manifold,''
  In *Arai, A. (ed.) et al.: A garden of quanta* 299-325.
  [hep-th/0302028].


\bibitem{Gauntlett:2000ib}
  J.~P.~Gauntlett, D.~Tong, P.~K.~Townsend,
  ``Multidomain walls in massive supersymmetric sigma models,''
  Phys.\ Rev.\  {\bf D64}, 025010 (2001).
  [hep-th/0012178];
  D.~Tong,
  ``The Moduli space of BPS domain walls,''
  Phys.\ Rev.\  {\bf D66}, 025013 (2002).
  [hep-th/0202012];
  Y.~Isozumi, K.~Ohashi and N.~Sakai,
  ``Exact wall solutions in 5-dimensional SUSY QED at finite coupling,''
  JHEP {\bf 0311}, 060 (2003)
  [arXiv:hep-th/0310189];
  M.~Eto, Y.~Isozumi, M.~Nitta, K.~Ohashi, K.~Ohta, N.~Sakai, Y.~Tachikawa,
  ``Global structure of moduli space for BPS walls,''
  Phys.\ Rev.\  {\bf D71}, 105009 (2005).
  [hep-th/0503033].

\bibitem{Bogomolny:1975de}
  E.~B.~Bogomolny,
  ``Stability of Classical Solutions,''
  Sov.\ J.\ Nucl.\ Phys.\  {\bf 24}, 449 (1976); 
  M.~K.~Prasad, C.~M.~Sommerfield,
  ``An Exact Classical Solution for the 't Hooft Monopole and the Julia-Zee Dyon,''
  Phys.\ Rev.\ Lett.\  {\bf 35}, 760-762 (1975).

\bibitem{Witten:1978mh}
  E.~Witten, D.~I.~Olive,
  ``Supersymmetry Algebras That Include Topological Charges,''
  Phys.\ Lett.\  {\bf B78}, 97 (1978).

\bibitem{Manton:1981mp}
  N.~S.~Manton,
  ``A Remark On The Scattering Of Bps Monopoles,''
  Phys.\ Lett.\  B {\bf 110} (1982) 54.

\bibitem{Atiyah:1985dv}
  M.~F.~Atiyah and N.~J.~Hitchin,
  ``Low-Energy Scattering Of Nonabelian Monopoles,''
  Phys.\ Lett.\  A {\bf 107}, 21 (1985); 
  M.~F.~Atiyah and N.~J.~Hitchin,
  ``The Geometry and Dynamics of Magnetic Monopoles. M.B. Porter Lectures,''
{\it  PRINCETON, USA: UNIV. PR. (1988) 133p}

\bibitem{Gibbons:1995yw}
  G.~W.~Gibbons and N.~S.~Manton,
  ``The Moduli space metric for well separated BPS monopoles,''
  Phys.\ Lett.\  B {\bf 356}, 32 (1995)
  [arXiv:hep-th/9506052].

\bibitem{Lee:1996kz}
  K.~M.~Lee, E.~J.~Weinberg and P.~Yi,
  ``The Moduli Space of Many BPS Monopoles for Arbitrary Gauge Groups,''
  Phys.\ Rev.\  D {\bf 54}, 1633 (1996)
  [arXiv:hep-th/9602167].

\bibitem{Taubes:1979tm}
  C.~H.~Taubes,
  ``Arbitrary N: Vortex Solutions To The First Order Landau-Ginzburg
  Equations,''
  Commun.\ Math.\ Phys.\  {\bf 72}, 277 (1980).

\bibitem{Ruback:1988ba}
  P.~J.~Ruback,
  ``Vortex String Motion In The Abelian Higgs Model,''
  Nucl.\ Phys.\  {\bf B296}, 669-678 (1988).

\bibitem{Shellard:1988zx}
  E.~P.~S.~Shellard, P.~J.~Ruback,
  ``Vortex Scattering In Two-dimensions,''
  Phys.\ Lett.\  {\bf B209}, 262-270 (1988).

\bibitem{Samols:1991ne}
  T.~M.~Samols,
  ``Vortex Scattering,''
  Commun.\ Math.\ Phys.\  {\bf 145}, 149 (1992).

\bibitem{Manton:2002wb}
  N.~S.~Manton and J.~M.~Speight,
  ``Asymptotic interactions of critically coupled vortices,''
  Commun.\ Math.\ Phys.\  {\bf 236}, 535 (2003)
  [arXiv:hep-th/0205307].

\bibitem{Chen:2004xu}
  H.~Y.~Chen and N.~S.~Manton,
  ``The Kaehler potential of Abelian Higgs vortices,''
  J.\ Math.\ Phys.\  {\bf 46}, 052305 (2005)
  [arXiv:hep-th/0407011].


\bibitem{Ward:1985ij}
  R.~S.~Ward,
  ``Slowly Moving Lumps In The Cp**1 Model In (2+1)-Dimensions,''
  Phys.\ Lett.\  B {\bf 158}, 424 (1985).

\bibitem{Eto:2006bb}
  M.~Eto, T.~Fujimori, T.~Nagashima, M.~Nitta, K.~Ohashi, N.~Sakai,
  ``Effective Action of Domain Wall Networks,''
  Phys.\ Rev.\  {\bf D75}, 045010 (2007).
  [hep-th/0612003];
  ``Dynamics of Domain Wall Networks,''
  Phys.\ Rev.\  {\bf D76}, 125025 (2007).
  [arXiv:0707.3267 [hep-th]].

\bibitem{Eto:2008mf}
  M.~Eto, T.~Fujimori, T.~Nagashima, M.~Nitta, K.~Ohashi, N.~Sakai,
  ``Dynamics of Strings between Walls,''
  Phys.\ Rev.\  {\bf D79}, 045015 (2009).
  [arXiv:0810.3495 [hep-th]].


\bibitem{Eto:2005cp}
  M.~Eto, Y.~Isozumi, M.~Nitta, K.~Ohashi and N.~Sakai,
  ``Webs of walls,''
  Phys.\ Rev.\  D {\bf 72}, 085004 (2005)
  [arXiv:hep-th/0506135];
  M.~Eto, Y.~Isozumi, M.~Nitta, K.~Ohashi and N.~Sakai,
  ``Non-abelian webs of walls,''
  Phys.\ Lett.\  B {\bf 632}, 384 (2006)
  [arXiv:hep-th/0508241];
  M.~Eto, Y.~Isozumi, M.~Nitta, K.~Ohashi, K.~Ohta and N.~Sakai,
  ``D-brane configurations for domain walls and their webs,''
  AIP Conf.\ Proc.\  {\bf 805}, 354 (2006)
  [arXiv:hep-th/0509127].

\bibitem{Eto:2005sw}
  M.~Eto, Y.~Isozumi, M.~Nitta and K.~Ohashi,
  ``1/2, 1/4 and 1/8 BPS equations in SUSY Yang-Mills-Higgs systems: Field
  theoretical brane configurations,''
  Nucl.\ Phys.\  B {\bf 752}, 140 (2006)
  [arXiv:hep-th/0506257].

\bibitem{Eto:2006pg}
  M.~Eto, Y.~Isozumi, M.~Nitta, K.~Ohashi and N.~Sakai,
  ``Solitons in the Higgs phase: The moduli matrix approach,''
  J.\ Phys.\ A  {\bf 39} (2006) R315
  [arXiv:hep-th/0602170].

\bibitem{Isozumi:2004vg}
  Y.~Isozumi, M.~Nitta, K.~Ohashi and N.~Sakai,
  ``All exact solutions of a 1/4 Bogomol'nyi-Prasad-Sommerfield equation,''
  Phys.\ Rev.\  D {\bf 71}, 065018 (2005)
  [arXiv:hep-th/0405129].

\bibitem{Witten:1976ck}
  E.~Witten,
  ``Some exact multipseudoparticle solutions of classical Yang-Mills  theory,''
  Phys.\ Rev.\ Lett.\  {\bf 38}, 121 (1977).

\bibitem{Krusch:2009tn}
  S.~Krusch and J.~M.~Speight,
  J.\ Math.\ Phys.\  {\bf 51}, 022304 (2010)
  [arXiv:0906.2007 [hep-th]].



\bibitem{Weinberg:1979er}
  E.~J.~Weinberg,
  ``Multivortex Solutions Of The Ginzburg-landau Equations,''
  Phys.\ Rev.\  {\bf D19}, 3008 (1979);
  ``Index Calculations for the Fermion-Vortex System,''
  Phys.\ Rev.\  {\bf D24}, 2669 (1981).


\bibitem{Isozumi:2004jc}
  Y.~Isozumi, M.~Nitta, K.~Ohashi, N.~Sakai,
  ``Construction of non-Abelian walls and their complete moduli space,''
  Phys.\ Rev.\ Lett.\  {\bf 93}, 161601 (2004).
  [hep-th/0404198];
  ``Non-Abelian walls in supersymmetric gauge theories,''
  Phys.\ Rev.\  {\bf D70}, 125014 (2004).
  [hep-th/0405194]; 
  M.~Eto, Y.~Isozumi, M.~Nitta, K.~Ohashi, K.~Ohta, N.~Sakai,
  ``D-brane construction for non-Abelian walls,''
  Phys.\ Rev.\  {\bf D71}, 125006 (2005).
  [hep-th/0412024];
  A.~Hanany, D.~Tong,
  ``On monopoles and domain walls,''
  Commun.\ Math.\ Phys.\  {\bf 266}, 647-663 (2006).
  [hep-th/0507140].


\bibitem{Goddard:1976qe}
  P.~Goddard, J.~Nuyts, D.~I.~Olive,
  ``Gauge Theories and Magnetic Charge,''
  Nucl.\ Phys.\  {\bf B125}, 1 (1977);
  C.~Montonen, D.~I.~Olive,
  ``Magnetic Monopoles as Gauge Particles?,''
  Phys.\ Lett.\  {\bf B72}, 117 (1977);
  E.~J.~Weinberg,
  ``Fundamental Monopoles and Multi-Monopole Solutions for Arbitrary Simple Gauge Groups,''
  Nucl.\ Phys.\  {\bf B167}, 500 (1980);
  R.~Auzzi, S.~Bolognesi, J.~Evslin, K.~Konishi, H.~Murayama,
  ``NonAbelian monopoles,''
  Nucl.\ Phys.\  {\bf B701}, 207-246 (2004).
  [hep-th/0405070];
  M.~Nitta and W.~Vinci,
  Nucl.\ Phys.\  B {\bf 848}, 121 (2011)
  [arXiv:1012.4057 [hep-th]].


\bibitem{Hanany:2003hp}
  A.~Hanany and D.~Tong,
  ``Vortices, instantons and branes,''
  JHEP {\bf 0307}, 037 (2003)
  [arXiv:hep-th/0306150];
  R.~Auzzi, S.~Bolognesi, J.~Evslin, K.~Konishi and A.~Yung,
  ``Nonabelian superconductors: Vortices and confinement in N = 2 SQCD,''
  Nucl.\ Phys.\  B {\bf 673}, 187 (2003)
  [arXiv:hep-th/0307287].

\bibitem{Shifman:2003uh}
  M.~Shifman, A.~Yung,
  ``Localization of nonAbelian gauge fields on domain walls at weak coupling (D-brane prototypes II),''
  Phys.\ Rev.\  {\bf D70}, 025013 (2004);
  [hep-th/0312257];
  M.~Eto, M.~Nitta, K.~Ohashi, D.~Tong,
  ``Skyrmions from instantons inside domain walls,''
  Phys.\ Rev.\ Lett.\  {\bf 95}, 252003 (2005).
  [hep-th/0508130]; 
  M.~Eto, T.~Fujimori, M.~Nitta, K.~Ohashi, N.~Sakai,
  ``Domain walls with non-Abelian clouds,''
  Phys.\ Rev.\  {\bf D77}, 125008 (2008).
  [arXiv:0802.3135 [hep-th]].


\bibitem{review}
  D.~Tong,
  ``TASI lectures on solitons,''
  arXiv:hep-th/0509216;
  ``Quantum Vortex Strings: A Review,''
  Annals Phys.\  {\bf 324}, 30 (2009)
  [arXiv:0809.5060 [hep-th]];
  K.~Konishi,
  ``The magnetic monopoles seventy-five years later,''
  Lect.\ Notes Phys.\  {\bf 737}, 471 (2008)
  [arXiv:hep-th/0702102];
  ``Advent of Non-Abelian Vortices and Monopoles-- further thoughts about
  duality and confinement,''
  Prog.\ Theor.\ Phys.\ Suppl.\  {\bf 177}, 83 (2009)
  [arXiv:0809.1370 [hep-th]];
  M.~Shifman and A.~Yung,
  ``Supersymmetric Solitons and How They Help Us Understand Non-Abelian   Gauge
  Theories,''
  Rev.\ Mod.\ Phys.\  {\bf 79}, 1139 (2007)
  [arXiv:hep-th/0703267]; 
an expanded version
in Cambridge University Press, 2009.


\bibitem{Eto:2005yh}
  M.~Eto, Y.~Isozumi, M.~Nitta, K.~Ohashi and N.~Sakai,
  ``Moduli space of non-Abelian vortices,''
  Phys.\ Rev.\ Lett.\  {\bf 96} (2006) 161601
  [arXiv:hep-th/0511088].

\bibitem{Eto:2006cx}
  M.~Eto, K.~Konishi, G.~Marmorini, M.~Nitta, K.~Ohashi, W.~Vinci and N.~Yokoi,
  ``Non-Abelian vortices of higher winding numbers,''
  Phys.\ Rev.\  D {\bf 74} (2006) 065021
  [arXiv:hep-th/0607070].

\bibitem{Eto:2006uw}
  M.~Eto, Y.~Isozumi, M.~Nitta, K.~Ohashi and N.~Sakai,
  ``Manifestly supersymmetric effective Lagrangians on BPS solitons,''
  Phys.\ Rev.\  D {\bf 73}, 125008 (2006)
  [arXiv:hep-th/0602289].

\bibitem{Eto:2006dx}
  M.~Eto {\it et al.},
  ``Non-Abelian duality from vortex moduli: a dual model of
  color-confinement,''
  Nucl.\ Phys.\  B {\bf 780}, 161 (2007)
  [arXiv:hep-th/0611313].

\bibitem{Eto:2010aj}
  M.~Eto, T.~Fujimori, S.~Bjarke Gudnason, Y.~Jiang, K.~Konishi, M.~Nitta, K.~Ohashi,
  ``Group Theory of Non-Abelian Vortices,''
  JHEP {\bf 1011}, 042 (2010).
  [arXiv:1009.4794 [hep-th]].

\bibitem{Eto:2006db}
  M.~Eto, K.~Hashimoto, G.~Marmorini, M.~Nitta, K.~Ohashi and W.~Vinci,
  ``Universal reconnection of non-Abelian cosmic strings,''
  Phys.\ Rev.\ Lett.\  {\bf 98} (2007) 091602
  [arXiv:hep-th/0609214].

\bibitem{Fujimori:2010fk}
  T.~Fujimori, G.~Marmorini, M.~Nitta, K.~Ohashi, N.~Sakai,
  ``The Moduli Space Metric for Well-Separated Non-Abelian Vortices,''
  Phys.\ Rev.\  {\bf D82}, 065005 (2010).
  [arXiv:1002.4580 [hep-th]].


\bibitem{Collie:2008za}
  B.~Collie,
  ``Dyonic Non-Abelian Vortices,''
  J.\ Phys.\ A  {\bf 42}, 085404 (2009)
  [arXiv:0809.0394 [hep-th]].



\bibitem{Shifman:2004dr}
  M.~Shifman and A.~Yung,
  ``Non-Abelian string junctions as confined monopoles,''
  Phys.\ Rev.\  D {\bf 70}, 045004 (2004)
  [arXiv:hep-th/0403149]; 
  A.~Hanany and D.~Tong,
  ``Vortex strings and four-dimensional gauge dynamics,''
  JHEP {\bf 0404}, 066 (2004)
  [arXiv:hep-th/0403158].

\bibitem{Eto:2004rz}
  M.~Eto, Y.~Isozumi, M.~Nitta, K.~Ohashi and N.~Sakai,
  ``Instantons in the Higgs phase,''
  Phys.\ Rev.\  D {\bf 72}, 025011 (2005)
  [arXiv:hep-th/0412048];
  T.~Fujimori, M.~Nitta, K.~Ohta, N.~Sakai, M.~Yamazaki,
  ``Intersecting Solitons, Amoeba and Tropical Geometry,''
  Phys.\ Rev.\  {\bf D78}, 105004 (2008).
  [arXiv:0805.1194 [hep-th]].



\bibitem{Eto:2009wq}
  M.~Eto, T.~Fujimori, T.~Nagashima, M.~Nitta, K.~Ohashi and N.~Sakai,
  ``Multiple Layer Structure of Non-Abelian Vortex,''
  Phys.\ Lett.\  B {\bf 678}, 254 (2009)
  [arXiv:0903.1518 [hep-th]].


\bibitem{Vachaspati:1991dz}
  T.~Vachaspati and A.~Achucarro,
  ``Semilocal cosmic strings,''
  Phys.\ Rev.\  D {\bf 44}, 3067 (1991);
  A.~Achucarro and T.~Vachaspati,
  ``Semilocal and electroweak strings,''
  Phys.\ Rept.\  {\bf 327}, 347 (2000)
  [Phys.\ Rept.\  {\bf 327}, 427 (2000)]
  [arXiv:hep-ph/9904229].

\bibitem{Leese:1992fn}
  R.~A.~Leese and T.~M.~Samols,
  ``Interaction of semilocal vortices,''
  Nucl.\ Phys.\  B {\bf 396}, 639 (1993).

\bibitem{Leese:1991hr}
  R.~A.~Leese,
  ``Q lumps and their interactions,''
  Nucl.\ Phys.\  B {\bf 366}, 283 (1991).


\bibitem{Eto:2007aw}
  M.~Eto, T.~Fujimori, M.~Nitta, K.~Ohashi, K.~Ohta and N.~Sakai,
  ``Statistical Mechanics of Vortices from D-branes and T-duality,''
  Nucl.\ Phys.\  B {\bf 788}, 120 (2008)
  [arXiv:hep-th/0703197].

\bibitem{Shifman:2006kd}
  M.~Shifman and A.~Yung,
  ``Non-Abelian semilocal strings in N = 2 supersymmetric QCD,''
  Phys.\ Rev.\  D {\bf 73}, 125012 (2006)
  [arXiv:hep-th/0603134].

\bibitem{Eto:2007yv}
  M.~Eto {\it et al.},
  ``On the moduli space of semilocal strings and lumps,''
  Phys.\ Rev.\  D {\bf 76} (2007) 105002
  [arXiv:0704.2218 [hep-th]].






\bibitem{Eto:2008yi}
  M.~Eto, T.~Fujimori, S.~B.~Gudnason, K.~Konishi, M.~Nitta, K.~Ohashi, W.~Vinci,
  ``Constructing Non-Abelian Vortices with Arbitrary Gauge Groups,''
  Phys.\ Lett.\  {\bf B669}, 98-101 (2008).
  [arXiv:0802.1020 [hep-th]].

\bibitem{Eto:2008qw}
  M.~Eto, T.~Fujimori, S.~B.~Gudnason, M.~Nitta, K.~Ohashi,
  ``SO and USp Kahler and Hyper-Kahler Quotients and Lumps,''
  Nucl.\ Phys.\  {\bf B815}, 495-538 (2009).
  [arXiv:0809.2014 [hep-th]];
  M.~Eto, T.~Fujimori, S.~B.~Gudnason, K.~Konishi, T.~Nagashima, M.~Nitta, K.~Ohashi, W.~Vinci,
  ``Non-Abelian Vortices in SO(N) and USp(N) Gauge Theories,''
  JHEP {\bf 0906}, 004 (2009).
  [arXiv:0903.4471 [hep-th]].

\bibitem{Ferretti:2007rp}
  L.~Ferretti, S.~B.~Gudnason, K.~Konishi,
  ``Non-Abelian vortices and monopoles in SO(N) theories,''
  Nucl.\ Phys.\  {\bf B789}, 84-110 (2008).
  [arXiv:0706.3854 [hep-th]];
  S.~B.~Gudnason, K.~Konishi,
  ``Low-energy U(1) x USp(2M) gauge theory from simple high-energy gauge group,''
  Phys.\ Rev.\  {\bf D81}, 105007 (2010).
  [arXiv:1002.0850 [hep-th]];
  S.~B.~Gudnason, Y.~Jiang, K.~Konishi,
  ``Non-Abelian vortex dynamics: Effective world-sheet action,''
  JHEP {\bf 1008}, 012 (2010).
  [arXiv:1007.2116 [hep-th]].

\bibitem{Eto:2006mz}
  M.~Eto, T.~Fujimori, Y.~Isozumi, M.~Nitta, K.~Ohashi, K.~Ohta and N.~Sakai,
  ``Non-Abelian vortices on cylinder: Duality between vortices and walls,''
  Phys.\ Rev.\  D {\bf 73}, 085008 (2006)
  [arXiv:hep-th/0601181].

\bibitem{Baptista:2008ex}
  J.~M.~Baptista,
  ``Non-abelian vortices on compact Riemann surfaces,''
  Commun.\ Math.\ Phys.\  {\bf 291}, 799 (2009)
  [arXiv:0810.3220 [hep-th]];
  A.~D.~Popov,
  ``Integrability of Vortex Equations on Riemann Surfaces,''
  Nucl.\ Phys.\  B {\bf 821}, 452 (2009)
  [arXiv:0712.1756 [hep-th]];
  ``Non-Abelian Vortices on Riemann Surfaces: an Integrable Case,''
  Lett.\ Math.\ Phys.\  {\bf 84}, 139 (2008)
  [arXiv:0801.0808 [hep-th]].

\bibitem{Baptista:2010rv}
  J.~M.~Baptista,
  ``On the $L^{2}$-metric of vortex moduli spaces,''
  Nucl.\ Phys.\  {\bf B844}, 308-333 (2011).
  [arXiv:1003.1296 [hep-th]].


\bibitem{Speight:1996px}
  J.~M.~Speight,
  ``Static intervortex forces,''
  Phys.\ Rev.\  D {\bf 55}, 3830 (1997)
  [arXiv:hep-th/9603155].

\bibitem{Auzzi:2007wj}
  R.~Auzzi, M.~Eto and W.~Vinci,
  ``Static Interactions of non-Abelian Vortices,''
  JHEP {\bf 0802}, 100 (2008)
  [arXiv:0711.0116 [hep-th]];
  R.~Auzzi, M.~Eto, S.~B.~Gudnason, K.~Konishi and W.~Vinci,
  ``On the Stability of Non-Abelian Semi-local Vortices,''
  Nucl.\ Phys.\  B {\bf 813}, 484 (2009)
  [arXiv:0810.5679 [hep-th]].



\bibitem{Collie:2008mx}
  B.~Collie and D.~Tong,
  ``The Dynamics of Chern-Simons Vortices,''
  Phys.\ Rev.\  D {\bf 78}, 065013 (2008)
  [arXiv:0805.0602 [hep-th]];
  S.~B.~Gudnason,
  ``Non-Abelian Chern-Simons vortices with generic gauge groups,''
  Nucl.\ Phys.\  B {\bf 821}, 151 (2009)
  [arXiv:0906.0021 [hep-th]].

\bibitem{Hashimoto:2005hi}
  K.~Hashimoto, D.~Tong,
  ``Reconnection of non-Abelian cosmic strings,''
  JCAP {\bf 0509}, 004 (2005).
  [hep-th/0506022].


\end{thebibliography}
\end{document}